\documentclass[11pt]{amsart}

\usepackage[utf8]{inputenc}
\usepackage[english]{babel}
\usepackage{amsmath,amssymb,amsthm,mathtools}
\usepackage{mathrsfs}
\usepackage{enumerate}
\usepackage{xcolor}
\usepackage{tikz-cd}
\usepackage[margin=2.8cm]{geometry}
\usepackage[colorlinks=true,linkcolor=blue!60!black,citecolor=green!40!black,urlcolor=blue!60!black]{hyperref}

\newtheorem{theorem}{Theorem}[section]
\newtheorem{lemma}[theorem]{Lemma}
\newtheorem{prop}[theorem]{Proposition}
\newtheorem{cor}[theorem]{Corollary}
\theoremstyle{definition}
\newtheorem{definition}[theorem]{Definition}
\newtheorem{assumption}[theorem]{Assumption}
\theoremstyle{remark}
\newtheorem{remark}[theorem]{Remark}
\newtheorem{example}[theorem]{Example}

\numberwithin{equation}{section}

\renewcommand{\epsilon}{\varepsilon}
\renewcommand{\Im}{\operatorname{Im}}
\renewcommand{\Re}{\operatorname{Re}}
\newcommand{\dd}{\,\mathrm{d}}
\newcommand{\ee}{\mathrm{e}}
\newcommand{\D}{\mathscr{D}}
\newcommand{\Hf}{\mathscr{H}}
\DeclareMathOperator{\sgn}{sgn}
\DeclareMathOperator{\Dom}{Dom}
\DeclareMathOperator{\Ran}{Ran}
\DeclareMathOperator{\Ker}{Ker}

\DeclareMathOperator{\diag}{diag}

\newcommand{\C}{\mathbb{C}}
\newcommand{\R}{\mathbb{R}}

\newcommand{\HS}{\mathrm{HS}}
\newcommand{\Zt}{\widetilde{Z}}
\newcommand{\Ah}{\widehat{A}}
\newcommand{\res}{\rho}

\newcommand{\ket}[1]{|#1\rangle}
\newcommand{\bra}[1]{\langle #1|}

\title[Non-local approximations of general point interactions]{Non-local approximations for the Schrödinger operator with a general point interaction via the non-relativistic limit}

\author{Luk\'{a}\v{s} Heriban}
\address{Department of Mathematics, Faculty of Nuclear Sciences and Physical Engineering, Czech Technical University in Prague, Trojanova 13, 120 00 Prague, Czechia}
\email{heribluk@fjfi.cvut.cz}

\date{\today}


\begin{document}

\begin{abstract}
We consider the one-dimensional Dirac operator $D_{A,c}$ with a general, not necessarily self-adjoint, point interaction at the origin, whose strength is described by an arbitrary complex $2\times 2$ matrix $A$ and suitably rescaled by the speed of light $c$, together with its non-local approximations $D_{A,c}^{\epsilon}$ obtained by replacing the Dirac distribution by a rank-one projection onto a scaled function $v_\epsilon$. It is known that $D_{A,c}^\epsilon \to D_{A,c}$ as $\epsilon\to 0+$ and that $D_{A,c}-c^2/2$ converges, as $c\to+\infty$, to the Schrödinger operator $H_A$ with the general point interaction, both in the norm resolvent sense. We prove that the first convergence is uniform with respect to $c$ in a neighbourhood of $+\infty$, and we provide explicit rates of convergence. Combining this with an explicit computation of the non-relativistic limit of $D_{A,c}^\epsilon$ for fixed $\epsilon$, we show that the two limits can be interchanged and that the corresponding double limit exists. As a by-product we obtain a family of non-local, rank-two approximations $H_A^\epsilon$ of $H_A$ that converge to $H_A$ in the norm resolvent sense. Contrary to the relativistic setting, the coupling constants of $H_A^\epsilon$ are renormalized; we give an explicit formula for this renormalization.
\end{abstract}

\maketitle

\section{Introduction}

Hamiltonians with point (zero-range) interactions are among the most popular exactly solvable models of quantum mechanics; see the monographs \cite{AGHH05, AlKu00} and references therein. In one dimension, the most general point interaction for the Schrödinger operator is described by four complex parameters. In the self-adjoint case, the corresponding operators form the four-real-parametric family of self-adjoint extensions of the restriction of $-\frac{\dd^2}{\dd x^2}$ to functions vanishing at the origin together with their derivatives \cite{Se86a, AlDaKu98, BrNi13}. Alternatively, they can be introduced starting from the formal expression
\begin{equation}\label{eq:formal-H}
\Hf_A = -\frac{\dd^2}{\dd x^2} + a_{11}\ket{\delta}\bra{\delta} + a_{12}\ket{\delta}\bra{\delta'} + a_{21}\ket{\delta'}\bra{\delta} + a_{22}\ket{\delta'}\bra{\delta'},\qquad A=(a_{ij})_{i,j=1}^2\in\C^{2,2},
\end{equation}
and the theory of distributions acting on discontinuous test functions \cite{Ku96, KuBo98, AlKu00}. The latter approach does not rely on the self-adjointness of the resulting operator and it was employed to treat non-self-adjoint point interactions in \cite{GrKu14}; see also \cite{HuKrSi15} for the more general setting of quantum graphs or \cite{GlKu21} and \cite{KuZn17} for the similar model of the non-self-adjoint Schrödinger operator with non-local point interactions.

The relativistic counterpart of these models is the one-dimensional Dirac operator with a point interaction. In the self-adjoint setting it was studied, e.g., in \cite{GeSe87, BeDa94, PaRi14}, and the general, not necessarily self-adjoint, case was treated in \cite{HeTu22}. The connection between both models is provided by the non-relativistic limit. In \cite{BeDa94} it was shown that, after a suitable $c$-dependent rescaling of the coupling parameters, the self-adjoint relativistic point interactions converge, as the speed of light $c$ tends to infinity, to the self-adjoint non-relativistic ones. This was extended to arbitrary complex coupling matrices in \cite[Thm.~5.1]{HeTu22}. See also \cite{KrKr24}, where the non-relativistic limit of the Dirac operator
on a half-line was studied, with the boundary condition rescaled with the
speed of light in the same way as in \eqref{eq:Ac}. For the free operators, the non-relativistic limit is classical, see \cite{GGT84} and \cite[Ch.~6]{Th92}.

Another natural question is how point interactions can be approximated by more regular potentials. For the Dirac operator, it is well known that approximations by local, scaled potentials $\epsilon^{-1}h(x/\epsilon)$ converge in the norm resolvent sense, but the coupling constant of the limit point interaction differs in general from the one suggested by the formal limit; this effect is called the \emph{renormalization of coupling constants} \cite{Se89, Hu97, Hu99, Tu20}. The same phenomenon appears for $\delta$-shell interactions in higher dimensions \cite{MaPi17, MaPi18, CaLoMaTu23}. On the other hand, Šeba \cite{Se89} observed for two particular self-adjoint couplings, and it was later proved in full generality in \cite[Thm.~4.1]{HeTu22}, that the \emph{non-local} approximations
\[
D_A^\epsilon = D_0 + A\otimes\ket{v_\epsilon}\bra{v_\epsilon},\qquad v_\epsilon(x) = \epsilon^{-1}v(\epsilon^{-1}x),
\]
converge to the point interaction $D_A$ without any renormalization. See also \cite{HeTu24} for similar result in the higher dimension. For the Schrödinger operator, the situation is different: already for the $\delta'$-interaction, natural non-local approximations require a renormalization \cite{Se86b}, and an approximation scheme avoiding the renormalization was proposed in \cite{AlNi00}, see also \cite{AlNi07}.

These facts lead to the following question. Introduce the speed of light $c$ into the relativistic models and denote by $D_{A,c}$ and $D_{A,c}^\epsilon$ the relativistic point interaction and its non-local approximation, respectively (see Section~\ref{sec:prelim} for the precise definitions and for the $c$-dependent scaling of $A$). Then we have the following picture
\begin{equation}\label{eq:diagram-intro}
\begin{tikzcd}[column sep=6em, row sep=4em]
D_{A,c}^\epsilon - \frac{c^2}{2} \arrow[r, "\epsilon\to 0+"] \arrow[d, dashed, "c\to+\infty"'] & D_{A,c} - \frac{c^2}{2} \arrow[d, "c\to+\infty"] \\
\textbf{?} \arrow[r, dashed, "\epsilon\to 0+"'] & H_A
\end{tikzcd}
\end{equation}
where the solid arrows denote the known norm resolvent convergences \cite[Thm.~4.1 and Thm.~5.1]{HeTu22}. It is natural to ask whether the diagram can be completed and whether it commutes. In other words, what is the non-relativistic limit of the non-local approximations and do these limits converge to $H_A$?

The main results of the present paper give a positive answer. Our approach is based on the following observation: although the resolvents of $D_{A,c}$ and $D_{A,c}^\epsilon$ contain matrices which blow up as $c\to+\infty$, they can be factorized in such a way that all $c$-dependent quantities remain bounded (Lemma~\ref{lem:factorization}). This allows us to prove our first main result.

\begin{itemize}
\item[(i)] (Theorem~\ref{thm:uniform}.) The convergence $D_{A,c}^\epsilon - c^2/2 \to D_{A,c} - c^2/2$ as $\epsilon\to 0+$ holds in the norm resolvent sense \emph{uniformly} with respect to $c\in [c_1,+\infty)$ for a sufficiently large $c_1$. If, in addition, $\int_\R|s|^{1/2}|v(s)|\dd s<\infty$, the rate of this convergence is $O(\epsilon^{1/2})$.
\end{itemize}

Next, we compute the non-relativistic limit of $D_{A,c}^\epsilon$ for every fixed $\epsilon>0$.

\begin{itemize}
\item[(ii)] (Theorems~\ref{thm:nonrel-eps} and~\ref{thm:identification}.) For every sufficiently small fixed $\epsilon>0$, $D_{A,c}^\epsilon - c^2/2$ converges in the norm resolvent sense, as $c\to+\infty$, to the Schrödinger operator with a rank-two non-local potential
\begin{equation}\label{eq:HAeps-intro}
H_A^\epsilon = H_0 + \sum_{j,k=1}^{2} (B_\epsilon)_{jk}\ket{u_j^\epsilon}\bra{u_k^\epsilon},\qquad u_1^\epsilon = v_\epsilon,\quad u_2^\epsilon = v_\epsilon',
\end{equation}
where the coupling matrix $B_\epsilon$ is given explicitly in \eqref{eq:Beps}. The rate of this convergence is $O(c^{-1})$.
\end{itemize}

Using the theory of pseudoresolvents, we show moreover that the non-relativistic limit of $D^\epsilon_{A,c}$ is the resolvent of a closed operator also if $v\notin H^1(\R)$, while for $v\in H^1(\R)$ and an exceptional value of $\epsilon$ it is merely the resolvent of a linear relation (Propositions~\ref{prop:Teps} and~\ref{prop:kernel-H1}).

Finally, the uniformity in (i) together with (ii) and the Moore--Osgood theorem on interchanging limits yield our second main result.

\begin{itemize}
\item[(iii)] (Theorem~\ref{thm:main}.) The diagram \eqref{eq:diagram-intro} completed by $H_A^\epsilon$ commutes, the double limit of $(D_{A,c}^\epsilon - c^2/2 - z)^{-1}$ as $(\epsilon,c)\to(0+,+\infty)$ exists, and, in particular, $H_A^\epsilon\to H_A$ in the norm resolvent sense as $\epsilon\to 0+$. Under the moment condition from (i), the corresponding rates are $O(\epsilon^{1/2})$ and $O(\epsilon^{1/2}+c^{-1})$, respectively.
\end{itemize}

It is possible to directly prove that the approximations $H_A^\epsilon$ converge in the norm resolvent sense to $H_A$. The proof given here is different and, we believe, more informative: it shows that the convergence $H_A^\epsilon\to H_A$ is inherited from the relativistic non-local approximations through the uniformity in (i). Interestingly, while no renormalization occurs along the upper branch of \eqref{eq:diagram-intro}, the coupling matrix $B_\epsilon$ in \eqref{eq:HAeps-intro} does differ from $A$; in fact, $B_\epsilon$ does not converge to $A$ unless $a_{22} = 0$ and $a_{12}a_{21}=0$. In Proposition~\ref{prop:naive} we describe the norm resolvent limits of the ``naive'' approximations $H_0 + \sum_{j,k} b_{jk}\ket{u_j^\epsilon}\bra{u_k^\epsilon}$ with an $\epsilon$-independent matrix $B$. For instance, if $b_{22}\neq 0$, then they converge to the $\delta$-interaction of strength $\det B/b_{22}$; in particular, the naive approximations of the $\delta'$-interaction converge to the free Hamiltonian, which is in accordance with \cite{Se86b}.

The paper is organized as follows. In Section~\ref{sec:prelim} we fix the notation and recall the definitions and basic properties of the operators $H_A$, $D_{A,c}$, and $D_{A,c}^\epsilon$. In Section~\ref{sec:factorization} we derive a factorization of the resolvents which is well adapted to the non-relativistic limit. Section~\ref{sec:uniform} contains the proof of the uniform convergence. In Section~\ref{sec:nonrel} we compute the non-relativistic limit of the non-local approximations, discuss it from the point of view of pseudoresolvents, and identify the limit operator. Section~\ref{sec:interchange} is devoted to the interchange of limits and to the discussion of the renormalization of coupling constants.

\section{Preliminaries}\label{sec:prelim}

\subsection{Notation and conventions}
We work in units in which $\hbar = 1$ and $2m=1$, i.e., the mass of the particle is $m = 1/2$ and the rest energy equals $mc^2 = c^2/2$. For $w\in\C\setminus[0,+\infty)$, $\sqrt{w}$ always denotes the branch of the square root with $\Im\sqrt{w}>0$. This branch is analytic on $\C\setminus[0,+\infty)$.

For a matrix $B\in\C^{n,k}$ we denote by $|B|$ its Frobenius (Hilbert--Schmidt) norm, $|B|^2 = \sum_{i,j}|B_{ij}|^2$. Recall that $|BC|\le |B||C|$, that $|UBU'|=|B|$ for unitary $U,U'$, that the operator norm of $B$ does not exceed $|B|$, and that for an invertible $X\in\C^{2,2}$ we have $X^{-1} = \operatorname{adj}(X)/\det X$ with $|\operatorname{adj}(X)| = |X|$, hence
\begin{equation}\label{eq:inverse-2x2}
|X^{-1}| = \frac{|X|}{|\det X|}.
\end{equation}

We denote by $\langle f,g\rangle = \int_\R\overline{f}g$ the inner product in $L^2(\R)$; it is linear in the second argument. For $f,g\in L^2(\R)$, $\ket{f}\bra{g}$ stands for the rank-one operator $\psi\mapsto\langle g,\psi\rangle f$. If $\psi=(\psi_1,\psi_2)^T\in L^2(\R;\C^2)$, then $\langle f,\psi\rangle := (\langle f,\psi_1\rangle,\langle f,\psi_2\rangle)^T$ and, for $B\in\C^{2,2}$, $B\otimes\ket{f}\bra{f}$ denotes the operator $\psi\mapsto f\,B\langle f,\psi\rangle$ on $L^2(\R;\C^2)\equiv L^2(\R)\otimes\C^2$.

An integral operator on $L^2(\R;\C^2)$ with a matrix-valued kernel $k\in L^2(\R^2;\C^{2,2})$ is Hilbert--Schmidt and $\|K\|\le\|K\|_\HS = \big(\int_{\R^2}|k(x,y)|^2\dd x\dd y\big)^{1/2}$.

The Pauli matrices and some other matrices used throughout the paper are
\[
\sigma_1=\begin{pmatrix}0&1\\1&0\end{pmatrix},\quad
\sigma_3=\begin{pmatrix}1&0\\0&-1\end{pmatrix},\quad
P_1 = E_{11} = \begin{pmatrix}1&0\\0&0\end{pmatrix},\quad
E_{22} = \begin{pmatrix}0&0\\0&1\end{pmatrix},\quad
W=\begin{pmatrix}1&0\\0&i\end{pmatrix}.
\]
For $A=(a_{ij})\in\C^{2,2}$ we put
\begin{equation}\label{eq:Ahat}
\Ah := WAW^* = \begin{pmatrix} a_{11} & -ia_{12}\\ ia_{21} & a_{22}\end{pmatrix},
\end{equation}
and note that $\det\Ah = \det A$ and $|\Ah| = |A|$. For an operator $T$ on $L^2(\R)$, $P_1\otimes T$ is the operator $\psi\mapsto (T\psi_1, 0)^T$ on $L^2(\R;\C^2)$.

For a function $f$ on $\R$ we put $(\tau_h f)(x) := f(x-h)$ and $\check f(x) := f(-x)$, and $u*f$ is the convolution. We will repeatedly use the following elementary facts.

\begin{lemma}\label{lem:elementary}
\begin{enumerate}[(a)]
\item Let $\varphi,\psi\in L^2(\R;\C^{2,2})$ and $X\in\C^{2,2}$. Then the integral operator with kernel $\varphi(x)X\psi(y)$ is Hilbert--Schmidt and its Hilbert--Schmidt norm does not exceed $\|\varphi\|_{2}|X|\|\psi\|_{2}$.
\item Let $k\in L^1(\R;\C^{2,2})$. Then the convolution operator $\psi\mapsto k*\psi$ is bounded on $L^2(\R;\C^2)$ and its norm does not exceed $\|k\|_1 := \int_\R|k(x)|\dd x$.
\item Let $u\in L^1(\R)$ and $f\in L^2(\R;\C^{2,2})$. Then $\|u*f\|_2\le\|u\|_1\|f\|_2$ and
\[
\Big\|u*f - \Big(\int_\R u\Big)f\Big\|_2 \le \int_\R |u(s)|\,\|\tau_s f - f\|_2\dd s.
\]
\end{enumerate}
\end{lemma}

\begin{proof}
(a) follows from $|\varphi(x)X\psi(y)|\le|\varphi(x)||X||\psi(y)|$ and the Fubini theorem. Since $(u*f)(x) - (\int u)f(x) = \int_\R u(s)\big(f(x-s)-f(x)\big)\dd s$, both (b) and (c) are consequences of the Minkowski integral inequality.
\end{proof}

Finally, we recall the Moore--Osgood theorem on interchanging limits. It is usually formulated for sequences of complex-valued functions, cf.~\cite[Thm.~7.11]{Ru76}, but the proof works verbatim in any Banach space. For the reader's convenience we include it.

\begin{lemma}[Moore--Osgood]\label{lem:moore-osgood}
Let $\mathcal X$ be a Banach space, $\epsilon_1>0$, $c_1>0$ and $f:(0,\epsilon_1)\times[c_1,+\infty)\to\mathcal X$. Assume that
\begin{enumerate}[(a)]
\item there is $g:[c_1,+\infty)\to\mathcal X$ such that $\lim_{\epsilon\to 0+}\sup_{c\ge c_1}\|f(\epsilon,c)-g(c)\|=0$;
\item for every $\epsilon\in(0,\epsilon_1)$, the limit $h(\epsilon):=\lim_{c\to+\infty}f(\epsilon,c)$ exists.
\end{enumerate}
Then both limits $\lim_{\epsilon\to 0+}h(\epsilon)$ and $\lim_{c\to+\infty}g(c)$ exist and are equal. Moreover, denoting this common value by $L$, for every $\eta>0$ there are $\epsilon'>0$ and $c'>0$ such that $\|f(\epsilon,c)-L\|<\eta$ for all $\epsilon\in(0,\epsilon')$ and $c>c'$.
\end{lemma}

\begin{proof}
Let $\eta>0$ and choose $\epsilon'>0$ such that $\|f(\epsilon,c)-g(c)\|<\eta$ for all $\epsilon\in(0,\epsilon')$ and all $c\ge c_1$. For $\epsilon,\tilde\epsilon\in(0,\epsilon')$ we get
\[
\|h(\epsilon)-h(\tilde\epsilon)\| = \lim_{c\to+\infty}\|f(\epsilon,c)-f(\tilde\epsilon,c)\|\le 2\eta.
\]
Hence $h$ satisfies the Cauchy condition as $\epsilon\to 0+$ and, by completeness, $L:=\lim_{\epsilon\to0+}h(\epsilon)$ exists; moreover, $\|h(\epsilon)-L\|\le 2\eta$ for $\epsilon\in(0,\epsilon')$. Fix such an $\epsilon$ and choose $c'$ such that $\|f(\epsilon,c)-h(\epsilon)\|<\eta$ for $c>c'$. Then
\[
\|g(c)-L\|\le\|g(c)-f(\epsilon,c)\|+\|f(\epsilon,c)-h(\epsilon)\|+\|h(\epsilon)-L\|<4\eta,\qquad c>c',
\]
which proves $\lim_{c\to+\infty}g(c) = L$. Finally, for $\epsilon\in(0,\epsilon')$ and $c>c'$ we have $\|f(\epsilon,c)-L\|\le\|f(\epsilon,c)-g(c)\|+\|g(c)-L\|<5\eta$.
\end{proof}

\subsection{Schrödinger operator with a general point interaction}

The one-dimensional free Schrödinger operator is
\[
\Dom H_0 = H^2(\R),\qquad H_0\psi = -\psi''.
\]
It is self-adjoint and $\sigma(H_0)=\sigma_{ac}(H_0)=[0,+\infty)$. For $z\in\C\setminus[0,+\infty)$, the resolvent $K_0(z):=(H_0-z)^{-1}$ is the integral operator with the kernel
\begin{equation}\label{eq:free-schr-res}
K_0(z;x,y) = \frac{i}{2\sqrt{z}}\,\ee^{i\sqrt{z}|x-y|}.
\end{equation}

We now give a meaning to the formal expression \eqref{eq:formal-H}. Following \cite{Ku96, KuBo98, AlKu00} (see also \cite{GrKu14} and \cite[Sect.~5.2]{HeTu22}), for $\psi\in H^2(\R\setminus\{0\}) = H^2(\R_-)\oplus H^2(\R_+)$ we extend the action of the distributions $\delta$ and $\delta'$ by
\[
(\delta,\psi) := \frac{\psi(0+)+\psi(0-)}{2},\qquad (\delta',\psi) := -\frac{\psi'(0+)+\psi'(0-)}{2},
\]
where $\psi(0\pm)$ and $\psi'(0\pm)$ are the one-sided traces at the origin. In the sense of distributions we have $\psi'' = \{\psi''\} + (\psi'(0+)-\psi'(0-))\delta + (\psi(0+)-\psi(0-))\delta'$, where $\{\psi''\}$ is the pointwise second derivative on $\R\setminus\{0\}$. Requiring that $\Hf_A\psi\in L^2(\R)$, i.e., that the singular contributions cancel out, we arrive at the transmission conditions
\begin{align*}
\psi'(0+)-\psi'(0-) &= a_{11}\frac{\psi(0+)+\psi(0-)}{2} - a_{12}\frac{\psi'(0+)+\psi'(0-)}{2},\\
\psi(0+)-\psi(0-) &= a_{21}\frac{\psi(0+)+\psi(0-)}{2} - a_{22}\frac{\psi'(0+)+\psi'(0-)}{2}.
\end{align*}
This motivates the following definition.

\begin{definition}\label{def:HA}
Let $A=(a_{ij})_{i,j=1}^2\in\C^{2,2}$. The Schrödinger operator with the point interaction of type $A$ is the operator
\begin{align*}
\Dom H_A &= \{\psi\in H^2(\R\setminus\{0\})\mid \Gamma_1\psi=A\Gamma_0\psi\},\\
(H_A\psi)(x) &= -\psi''(x),\qquad x\in\R\setminus\{0\},
\end{align*}
where $\Gamma_0,\Gamma_1: H^2(\R\setminus\{0\})\to\C^2$ are given by
\[
\Gamma_0\psi = \frac{1}{2}\begin{pmatrix}
\psi(0+)+\psi(0-)\\
-\psi'(0+)-\psi'(0-)
\end{pmatrix},\qquad \Gamma_1\psi = \begin{pmatrix}
\psi'(0+)-\psi'(0-)\\
\psi(0+)-\psi(0-)
\end{pmatrix}.
\]
\end{definition}

\begin{example}
For $A = a E_{11}$ we obtain the $\delta$-interaction of strength $a$: $\psi$ is continuous at $0$ and $\psi'(0+)-\psi'(0-) = a\psi(0)$. For $A=bE_{22}$ we obtain the $\delta'$-interaction: $\psi'$ is continuous at $0$ and $\psi(0+)-\psi(0-) = -b\psi'(0)$.
\end{example}

\begin{remark}\label{rem:param}
In \cite[Sect.~5.2]{HeTu22} the transmission condition was written as $\widetilde\Gamma_1\psi = V A V^*\widetilde\Gamma_2\psi$ with $V=\diag(i,1)$, $\widetilde\Gamma_1=\Gamma_1$, $\widetilde\Gamma_2=\Gamma_0$ and with the Hamiltonian multiplied by $1/(2m)$. Since $V^* A V = WAW^*$, the operator $H_A$ of Definition~\ref{def:HA} coincides with the operator $H_{\Ah}$ of \cite{HeTu22} for $m = 1/2$, where $\Ah$ is given by \eqref{eq:Ahat}.
\end{remark}

We summarize the basic properties of $H_A$, see \cite{GrKu14} and \cite[Appendix]{HeTu22}. The operator $H_A$ is closed, $(H_A)^* = H_{A^*}$, and $H_A$ is self-adjoint if and only if $A$ is hermitian. Put
\begin{equation}\label{eq:Mz}
\Zt(z) := \begin{pmatrix}
z^{-1/2} & 0\\
0 & z^{1/2}
\end{pmatrix},\qquad M_z := I+\frac{i}{2}A\Zt(z),\qquad z\in\C\setminus[0,+\infty),
\end{equation}
where $z^{1/2} := \sqrt{z}$ and $z^{-1/2}:=1/\sqrt z$.

\begin{prop}[{\cite[Prop.~6.2 and Rem.~6.3]{HeTu22}, \cite{GrKu14}}]\label{prop:HA}
Let $z\in\C\setminus[0,+\infty)$. Then $z\in\res(H_A)$ if and only if
\[
\det M_z = 1 + \frac{i}{2}\Big(\frac{a_{11}}{\sqrt z} + a_{22}\sqrt{z}\Big) - \frac{\det A}{4} \neq 0,
\]
and otherwise $z$ is an eigenvalue of $H_A$. If $z\in\res(H_A)$, then $(H_A-z)^{-1}$ is the integral operator with the kernel
\begin{equation}\label{eq:res-HA}
(H_A-z)^{-1}(x,y)=\frac{i}{2\sqrt{z}}\ee^{i\sqrt{z}|x-y|}
+\frac{1}{4}\begin{pmatrix}
z^{-1/2} & i\sgn(x)
\end{pmatrix}M_z^{-1}A\begin{pmatrix}
z^{-1/2} \\
i\sgn(y)
\end{pmatrix}\ee^{i\sqrt{z}(|x|+|y|)}.
\end{equation}
\end{prop}

Since $(H_A-z)^{-1}-K_0(z)$ is of rank at most two, the Weyl theorem \cite[Thm.~XIII.14]{ReSi78} yields $\sigma_{ess}(H_A) = [0,+\infty)$ whenever $\res(H_A)\neq\emptyset$.

\begin{remark}
Formula \eqref{eq:res-HA} is \cite[eq.~(57)]{HeTu22} rewritten in the parametrization of Definition~\ref{def:HA}, cf.~Remark~\ref{rem:param}. Note that, formally, $K_0(z)\delta = K_0(z;\cdot,0)$ and $K_0(z)\delta' = -\frac12\sgn(\cdot)\ee^{i\sqrt z|\cdot|}$, so \eqref{eq:res-HA} may be written in the suggestive Krein-type form
\[
(H_A-z)^{-1} = K_0(z) - \Phi_0(x) M_z^{-1} A\, \Phi_0(y)^T,\qquad \Phi_0 := \big(K_0(z)\delta, K_0(z)\delta'\big).
\]
Its regularized version \eqref{eq:Leps} below will appear naturally in the non-relativistic limit of the non-local approximations.
\end{remark}

\subsection{Dirac operator with a general point interaction}

For $c>0$, the one-dimensional free Dirac operator of a particle of mass $m = 1/2$ is
\[
\Dom D_{0,c} = H^1(\R;\C^2),\qquad D_{0,c}\psi = -ic\,\sigma_1\psi' + \frac{c^2}{2}\sigma_3\psi.
\]
It is self-adjoint with $\sigma(D_{0,c}) = \sigma_{ac}(D_{0,c}) = (-\infty,-c^2/2]\cup[c^2/2,+\infty)$. For $w\in\res(D_{0,c})$, the resolvent $R_c(w):=(D_{0,c}-w)^{-1}$ is the integral operator with the kernel, cf.~\cite[Sect.~5.1]{HeTu22},
\begin{equation}\label{eq:free-dirac-res}
R_c(w;x,y) = \frac{i}{2c}\big(Z_c(w) + \sgn(x-y)\sigma_1\big)\ee^{ik_c(w)|x-y|},
\end{equation}
where
\[
Z_c(w) = \begin{pmatrix}\zeta_c(w) & 0\\ 0 & \zeta_c(w)^{-1}\end{pmatrix},\qquad \zeta_c(w) = \frac{w+c^2/2}{ck_c(w)},\qquad ck_c(w) = \sqrt{w^2-c^4/4}.
\]

Let $A\in\C^{2,2}$ and put
\begin{equation}\label{eq:Ac}
A_c := C A C^*,\qquad C := \begin{pmatrix}c^{-1/2} & 0\\ 0 & ic^{1/2}\end{pmatrix},\qquad\text{i.e.,}\qquad A_c = \begin{pmatrix} c^{-1}a_{11} & -ia_{12}\\ ia_{21} & c\,a_{22}\end{pmatrix}.
\end{equation}
Note that $C = S_c W$ with $S_c := \diag(c^{-1/2},c^{1/2})$, so that $A_c = S_c\Ah S_c$ and $\det A_c = \det A$. The Dirac operator with the general point interaction is formally given by
\[
\D_{A,c} = -ic\,\sigma_1\frac{\dd}{\dd x} + \frac{c^2}{2}\sigma_3 + c\,A_c\otimes\ket{\delta}\bra{\delta}.
\]
Extending the action of $\delta$ to functions from $H^1(\R\setminus\{0\};\C^2)$ as above and requiring the cancellation of the singular parts, we get the transmission condition $-ic\sigma_1(\psi(0+)-\psi(0-)) + cA_c\frac{\psi(0+)+\psi(0-)}{2}=0$, see \cite[Sect.~1]{HeTu22}. Hence we define
\begin{align*}
\Dom D_{A,c} &= \left\{\psi\in H^1(\R \setminus \{0\}; \C^2) \,\middle|\, (2i\sigma_1 - A_c)\psi(0+) = (2i\sigma_1 + A_c)\psi(0-) \right\}, \\
(D_{A,c}\psi)(x) &= -ic\,\sigma_1 \psi'(x) + \frac{c^2}{2} \sigma_3 \psi(x), \qquad x\in \R\setminus\{0\}.
\end{align*}
The scaling \eqref{eq:Ac} of the entries of $A$ is the one which leads to a non-trivial non-relativistic limit, see \cite{BeDa94} and \cite[Sect.~5.2]{HeTu22}.

Denote by $D^{(\mu)}_B$ the Dirac operator with mass $\mu$, speed of light equal to one and point interaction given by $B\in\C^{2,2}$ as introduced in \cite{HeTu22}. We have $D_{A,c} = c\,D^{(c/2)}_{A_c}$ and hence $(D_{A,c}-w)^{-1} = c^{-1}(D^{(c/2)}_{A_c}-w/c)^{-1}$. Consequently, \cite[Prop.~2.1, Prop.~2.3, Cor.~2.4, Thm.~3.2 and Thm.~3.6]{HeTu22} yield the following statement, see also \cite[Sect.~5.1]{HeTu22}.

\begin{prop}\label{prop:DA}
The operator $D_{A,c}$ is closed, $(D_{A,c})^* = D_{A^*,c}$, and $D_{A,c}$ is self-adjoint if and only if $A$ is hermitian. Let $w\in\res(D_{0,c})$. Then $w\in\res(D_{A,c})$ if and only if $\det\big(I+\frac{i}{2}A_cZ_c(w)\big)\neq 0$, and in that case
\begin{equation}\label{eq:res-DA}
(D_{A,c}-w)^{-1}(x,y) = R_c(w;x,y) - c\,R_c(w;x,0)\Big(I+\frac{i}{2}A_cZ_c(w)\Big)^{-1}A_c\,R_c(w;0,y).
\end{equation}
\end{prop}

The non-relativistic limit of $D_{A,c}$ was established in \cite{HeTu22}.

\begin{theorem}[{\cite[Thm.~5.1]{HeTu22}}]\label{thm:HeTu-nonrel}
Let $z\in\res(H_A)\setminus[0,+\infty)$. Then $z+c^2/2\in\res(D_{A,c})$ for all sufficiently large $c$ and
\begin{equation}\label{eq:nonrel-point}
\lim_{c\to+\infty}\Big\| \Big(D_{A,c} - \frac{c^2}{2} - z\Big)^{-1} - P_1\otimes(H_A - z)^{-1} \Big\| = 0.
\end{equation}
\end{theorem}

\subsection{Non-local approximations}

Throughout the paper, the function $v$ satisfies the following assumption.

\begin{assumption}\label{ass:v}
$v\in L^1(\R;\R)\cap L^2(\R;\R)$ and $\int_\R v = 1$.
\end{assumption}

For $\epsilon>0$ put $v_\epsilon(x) := \epsilon^{-1}v(\epsilon^{-1}x)$. Then $\int_\R v_\epsilon = 1$, $\|v_\epsilon\|_1=\|v\|_1$, $n_\epsilon:=\|v_\epsilon\|_2^2 = \epsilon^{-1}\|v\|_2^2$, and $v_\epsilon\to\delta$ in the sense of distributions as $\epsilon\to 0+$. Since, formally, $\ket{v_\epsilon}\bra{v_\epsilon}\to\ket{\delta}\bra{\delta}$, the natural candidate for a regular approximation of $D_{A,c}$ is
\[
\Dom D_{A,c}^\epsilon = H^1(\R;\C^2),\qquad D_{A,c}^\epsilon = D_{0,c} + c\,A_c\otimes\ket{v_\epsilon}\bra{v_\epsilon}.
\]
The perturbation is bounded, so $D^\epsilon_{A,c}$ is closed. We will also need the function
\begin{equation}\label{eq:alpha1}
\alpha_1(w) := \int_{\R^2} v(x)\,\ee^{iw|x-y|}\,v(y)\dd x\dd y,\qquad \Im w\ge 0,
\end{equation}
which is continuous on $\{w\in\C\mid\Im w\ge 0\}$ by the dominated convergence theorem and satisfies $\alpha_1(0)=1$, $|\alpha_1(w)|\le\|v\|_1^2$. Note that $\int_{\R^2}v_\epsilon(x)\ee^{ik|x-y|}v_\epsilon(y)\dd x\dd y = \alpha_1(\epsilon k)$. To quantify the rates of convergence we will use the moment
\begin{equation}\label{eq:moment}
m_v := \int_\R|s|^{1/2}|v(s)|\dd s\in[0,+\infty],
\end{equation}
which is finite, e.g., if $v$ is compactly supported or if $\int_\R|s||v(s)|\dd s<\infty$.

Since $D_{A,c}^\epsilon = c\,(D^{(c/2)}_{0} + A_c\otimes\ket{v_\epsilon}\bra{v_\epsilon})$, \cite[Prop.~4.4 and Thm.~4.1]{HeTu22} give the following.

\begin{prop}\label{prop:DAeps}
Let $\epsilon>0$ and $w\in\res(D_{0,c})$ be such that $\det\big(I+\frac{i}{2}\alpha_1(\epsilon k_c(w))A_cZ_c(w)\big)\neq0$. Then $w\in\res(D^\epsilon_{A,c})$ and $(D^\epsilon_{A,c}-w)^{-1}$ is the integral operator with the kernel
\begin{multline}\label{eq:res-DAeps}
(D^\epsilon_{A,c}-w)^{-1}(x,y) = R_c(w;x,y)\\ - c\int_{\R^2} R_c(w;x,\epsilon s)v(s)\Big(I+\frac{i}{2}\alpha_1(\epsilon k_c(w))A_cZ_c(w)\Big)^{-1}A_c\,v(t)R_c(w;\epsilon t,y)\dd s\dd t.
\end{multline}
Moreover, for every fixed $c>0$ and $w\in\res(D_{A,c})$,
\begin{equation}\label{eq:rel-nonlocal-limit}
\lim_{\epsilon\to0+}\big\|(D_{A,c}^\epsilon-w)^{-1}-(D_{A,c}-w)^{-1}\big\| = 0.
\end{equation}
\end{prop}

The proof of \eqref{eq:rel-nonlocal-limit} in \cite{HeTu22} uses estimates whose constants depend on $|Z_c(w)|$, which grows like $c$ for $w=z+c^2/2$. Hence it does not give any information on the behaviour of the convergence as $c\to+\infty$. The aim of the next two sections is to show that, after the shift by the rest energy, the convergence is in fact uniform in $c$.

\section{Factorization of the resolvents near the upper threshold}\label{sec:factorization}

Throughout this section, $z\in\C\setminus[0,+\infty)$ is fixed and we put
\[
w_c := z+\frac{c^2}{2},\qquad \lambda_c := z+\frac{z^2}{c^2}.
\]

\begin{lemma}\label{lem:asymptotics}
There is $c_z>0$ such that for all $c\ge c_z$ we have $\lambda_c\in\C\setminus[0,+\infty)$, $w_c\in\res(D_{0,c})$, and
\[
\kappa_c := k_c(w_c) = \sqrt{\lambda_c},\qquad \zeta_c(w_c) = c\,\tau_c,\qquad\text{where}\quad \tau_c:=\frac{1+z/c^2}{\kappa_c}.
\]
Moreover, $\kappa_c\to\sqrt z$ and $\tau_c\to z^{-1/2}$ as $c\to+\infty$.
\end{lemma}

\begin{proof}
We have $w_c^2-c^4/4 = z^2+c^2 z = c^2\lambda_c$. Since $\lambda_c\to z$ and $[0,+\infty)$ is closed, $\lambda_c\notin[0,+\infty)$ for $c$ large enough. A point $w\in\C$ belongs to $\sigma(D_{0,c})$ if and only if $w^2-c^4/4\in[0,+\infty)$; hence $w_c\in\res(D_{0,c})$. If $r$ is the square root of $c^2\lambda_c$ with $\Im r>0$, then $r/c$ is the square root of $\lambda_c$ with positive imaginary part, which yields $ck_c(w_c) = c\sqrt{\lambda_c}$. Furthermore, $\zeta_c(w_c) = (z+c^2)/(c\sqrt{\lambda_c}) = c\tau_c$. The limits follow from the continuity of the square root on $\C\setminus[0,+\infty)$.
\end{proof}

Since $\Im\sqrt{z}>0$, the number $z^{-1/2} = \overline{\sqrt z}/|z|$ has a negative imaginary part. Enlarging $c_z$ if necessary, we may thus assume that $\Im\tau_c<0$ for $c\ge c_z$. We denote by $\vartheta_c := \tau_c^{1/2}$ and $\vartheta_\infty := (z^{-1/2})^{1/2}$ the principal square roots; then $\vartheta_c\to\vartheta_\infty$, $\vartheta_\infty^2 = z^{-1/2}$, and $\vartheta_\infty^{-2} = z^{1/2}$. For $c\in[c_z,+\infty]$ we introduce
\begin{equation}\label{eq:def-Pi}
\Pi_c := \begin{pmatrix}\vartheta_c & 0\\ 0 & c^{-1}\vartheta_c^{-1}\end{pmatrix},\qquad
\Delta_c := \begin{pmatrix}\vartheta_c & 0\\ 0 & \vartheta_c^{-1}\end{pmatrix},\qquad
F_c(x) := \big(I+\sgn(x)\sigma_1\big)\ee^{i\kappa_c|x|},
\end{equation}
\begin{equation}\label{eq:def-Q}
Q_c := \Delta_c\Ah\Delta_c,\qquad N_c(\alpha) := \Big(I+\frac{i}{2}\alpha Q_c\Big)^{-1}Q_c,\qquad d_c(\alpha) := \det\Big(I+\frac{i}{2}\alpha Q_c\Big),
\end{equation}
where $\kappa_\infty := \sqrt z$, $c^{-1}=0$ for $c=+\infty$, $\alpha\in\C$, and $N_c(\alpha)$ is defined whenever $d_c(\alpha)\neq0$. Thus $\Pi_\infty = \diag(\vartheta_\infty,0)$ and $\Delta_\infty^2 = \Zt(z)$. More generally, $\Delta_c^2 = \Zt_c := \diag(\tau_c,\tau_c^{-1})$ and, since $W$ is diagonal and unitary,
\begin{equation}\label{eq:dc}
d_c(\alpha) = \det\Big(I + \frac{i}{2}\alpha A\Zt_c\Big) = 1 + \frac{i}{2}\alpha\big(a_{11}\tau_c + a_{22}\tau_c^{-1}\big) - \frac{\alpha^2}{4}\det A.
\end{equation}
In particular, $d_\infty(1) = \det M_z$. Note also that $\Pi_c\to\Pi_\infty$, $\Delta_c\to\Delta_\infty$ and $Q_c\to Q_\infty$ as $c\to+\infty$.

The following lemma is the key observation of the paper. It shows that the resolvents of $D_{A,c}$ and $D^\epsilon_{A,c}$ at $w_c$ can be written in terms of the quantities \eqref{eq:def-Pi}, \eqref{eq:def-Q}, which, contrary to $Z_c(w_c)$, remain bounded as $c\to+\infty$.

\begin{lemma}\label{lem:factorization}
Let $c\ge c_z$ and $\epsilon>0$.
\begin{enumerate}[(a)]
\item $R_c(w_c;x,y) = \frac{i}{2}\Pi_cF_c(x-y)\Pi_c$.
\item If $d_c(1)\neq0$, then $w_c\in\res(D_{A,c})$ and
\[
(D_{A,c}-w_c)^{-1}(x,y) - R_c(w_c;x,y) = \frac14\Pi_cF_c(x)N_c(1)F_c(-y)\Pi_c.
\]
\item If $d_c(\alpha_1(\epsilon\kappa_c))\neq0$, then $w_c\in\res(D^\epsilon_{A,c})$ and
\[
(D^\epsilon_{A,c}-w_c)^{-1}(x,y) - R_c(w_c;x,y) = \frac14\Pi_c\,(v_\epsilon*F_c)(x)\,N_c\big(\alpha_1(\epsilon\kappa_c)\big)\,(v_\epsilon*\check F_c)(y)\,\Pi_c.
\]
\end{enumerate}
\end{lemma}

\begin{proof}
Put $P_c:=c^{1/2}\Pi_c = \diag(c^{1/2}\vartheta_c, c^{-1/2}\vartheta_c^{-1})$. By Lemma~\ref{lem:asymptotics}, $P_c^2 = \diag(c\tau_c,(c\tau_c)^{-1}) = Z_c(w_c)$. Moreover, $P_c^{-1}\sigma_1P_c^{-1}=\sigma_1$, whence
\[
Z_c(w_c)\pm\sigma_1 = P_c(I\pm\sigma_1)P_c = c\,\Pi_c(I\pm\sigma_1)\Pi_c.
\]
Inserting this into \eqref{eq:free-dirac-res} we get (a).

Next, since $P_cS_c = \Delta_c$, we have $P_cA_cP_c = P_cS_c\Ah S_cP_c = Q_c$. Let $\alpha\in\C$. By the Sylvester determinant identity, $\det(I+\frac{i}{2}\alpha A_cZ_c(w_c)) = \det(I+\frac{i}{2}\alpha A_cP_c^2) = \det(I+\frac{i}{2}\alpha P_cA_cP_c) = d_c(\alpha)$. If $d_c(\alpha)\neq0$, then from $P_c(I+\frac{i}{2}\alpha A_cP_c^2) = (I+\frac{i}{2}\alpha Q_c)P_c$ we infer
\begin{equation}\label{eq:pushthrough}
P_c\Big(I+\frac{i}{2}\alpha A_cZ_c(w_c)\Big)^{-1}A_cP_c = \Big(I+\frac{i}{2}\alpha Q_c\Big)^{-1}Q_c = N_c(\alpha).
\end{equation}
Using (a) and \eqref{eq:pushthrough}, we get for all $x,y,s,t\in\R$
\begin{align*}
&c\,R_c(w_c;x,s)\Big(I+\frac{i}{2}\alpha A_cZ_c(w_c)\Big)^{-1}A_cR_c(w_c;t,y)\\
&\qquad= -\frac{c}{4}\Pi_cF_c(x-s)\,\Pi_c\Big(I+\frac{i}{2}\alpha A_cZ_c(w_c)\Big)^{-1}A_c\Pi_c\,F_c(t-y)\Pi_c\\
&\qquad= -\frac{1}{4}\Pi_cF_c(x-s)N_c(\alpha)F_c(t-y)\Pi_c.
\end{align*}
With $\alpha=1$ and $s=t=0$, Proposition~\ref{prop:DA} yields (b). With $\alpha=\alpha_1(\epsilon\kappa_c)$, $s$ replaced by $\epsilon s$ and $t$ by $\epsilon t$, Proposition~\ref{prop:DAeps} yields (c), because
\[
\int_\R v(s)F_c(x-\epsilon s)\dd s = (v_\epsilon*F_c)(x),\qquad \int_\R v(t)F_c(\epsilon t-y)\dd t = \int_\R v_\epsilon(u)\check F_c(y-u)\dd u = (v_\epsilon*\check F_c)(y).\qedhere
\]
\end{proof}

At $c=+\infty$, the same expressions reproduce the resolvent of the Schrödinger operator.

\begin{lemma}\label{lem:c-infinity}
Let $z\in\C\setminus[0,+\infty)$. Then
\begin{enumerate}[(a)]
\item $\frac{i}{2}\Pi_\infty F_\infty(x-y)\Pi_\infty = K_0(z;x,y)\,P_1$;
\item if $\alpha\in\C$ and $d_\infty(\alpha)\neq 0$, then $N_\infty(\alpha) = \Delta_\infty W\big(I+\frac{i}{2}\alpha A\Zt(z)\big)^{-1}AW^*\Delta_\infty$;
\item if $\det M_z\neq0$, then $\frac14\Pi_\infty F_\infty(x)N_\infty(1)F_\infty(-y)\Pi_\infty = \big[(H_A-z)^{-1}(x,y)-K_0(z;x,y)\big]P_1$.
\end{enumerate}
\end{lemma}

\begin{proof}
For any $X\in\C^{2,2}$ we have $\Pi_\infty X\Pi_\infty = \vartheta_\infty^2X_{11}P_1 = z^{-1/2}X_{11}P_1$, and $[F_\infty(x)]_{11} = \ee^{i\sqrt z|x|}$; this gives (a). Since $Q_\infty = (\Delta_\infty W)A(W^*\Delta_\infty)$ and $W^*\Delta_\infty^2W = \Zt(z)$, (b) follows from the push-through identity $(I+XY)^{-1}X = X(I+YX)^{-1}$ with $X=\Delta_\infty W$ and $Y=\frac{i}{2}\alpha AW^*\Delta_\infty$. To prove (c), note that the first row of $F_\infty(x)$ is $(1,\sgn x)\ee^{i\sqrt z|x|}$ and the first column of $F_\infty(-y)$ is $(1,-\sgn y)^T\ee^{i\sqrt z|y|}$. Using (b) we get
\[
(1,\sgn x)\Delta_\infty W = \vartheta_\infty^{-1}\big(z^{-1/2},\, i\sgn x\big),\qquad W^*\Delta_\infty\begin{pmatrix}1\\-\sgn y\end{pmatrix} = \vartheta_\infty^{-1}\begin{pmatrix}z^{-1/2}\\ i\sgn y\end{pmatrix}.
\]
Consequently,
\[
\frac14\big[\Pi_\infty F_\infty(x)N_\infty(1)F_\infty(-y)\Pi_\infty\big]_{11} = \frac14\big(z^{-1/2}, i\sgn x\big)M_z^{-1}A\begin{pmatrix}z^{-1/2}\\ i\sgn y\end{pmatrix}\ee^{i\sqrt z(|x|+|y|)},
\]
which is the second term of \eqref{eq:res-HA}. Here we used $\Pi_\infty X\Pi_\infty = \vartheta_\infty^2X_{11}P_1$ and $\vartheta_\infty^2\vartheta_\infty^{-2}=1$.
\end{proof}

Next we collect the uniform bounds which will be used below. Let us put
\[
\gamma := \frac{1}{2}\Im\sqrt{z}>0,\qquad K:=2|\sqrt z|,\qquad M_0 := 2\big(|z|^{1/2}+|z|^{-1/2}\big).
\]

\begin{lemma}\label{lem:uniform-bounds}
There is $c_0\ge\max\{c_z,1\}$ such that for all $c\in[c_0,+\infty]$ the following holds.
\begin{enumerate}[(a)]
\item $\Im\kappa_c\ge\gamma$, $|\kappa_c|\le K$, $|\tau_c|+|\tau_c|^{-1}\le M_0$. Consequently, $|\Pi_c|^2\le M_0$, $|\Delta_c|^2\le M_0$ and $|Q_c|\le M_0|A|$.
\item $|F_c(x)|\le 2\ee^{-\gamma|x|}$ for all $x\in\R$; in particular $\|F_c\|_2 = \|\check F_c\|_2\le 2\gamma^{-1/2}$.
\item For all $h\in\R$,
\[
\|\tau_hF_c-F_c\|_2\le\omega(h) := \min\Big\{4\gamma^{-1/2},\ \big(8K^2h^2\gamma^{-1}\ee^{2\gamma|h|}+16|h|\big)^{1/2}\Big\},
\]
and the same estimate holds for $\check F_c$.
\item For all $\epsilon>0$, $|\alpha_1(\epsilon\kappa_c)-1|\le\eta(\epsilon)$, where
\[
\eta(\epsilon) := \int_{\R^2}|v(x)||v(y)|\min\{2,\,K\epsilon|x-y|\}\dd x\dd y\xrightarrow{\epsilon\to0+}0.
\]
\item $F_c\to F_\infty$ in $L^2(\R;\C^{2,2})$ as $c\to+\infty$.
\item There is $C_\omega>0$ such that $\omega(h)\le C_\omega|h|^{1/2}$ for all $h\in\R$, and $\eta(\epsilon)\le 2(2K\epsilon)^{1/2}\|v\|_1m_v$ for all $\epsilon>0$.
\item There is $C>0$ such that for all $c\in[c_0,+\infty)$ and $\epsilon>0$,
\begin{gather*}
|\kappa_c-\sqrt z|+|Q_c-Q_\infty|+\|F_c-F_\infty\|_1+\|F_c-F_\infty\|_2\le Cc^{-2},\qquad |\Pi_c-\Pi_\infty|\le Cc^{-1},\\
|\alpha_1(\epsilon\kappa_c)-\alpha_1(\epsilon\sqrt z)|\le\frac{\|v\|_1^2}{\ee\gamma}|\kappa_c-\sqrt z|.
\end{gather*}
\end{enumerate}
\end{lemma}

\begin{proof}
(a) The first three inequalities hold for large $c$ since $\kappa_c\to\sqrt z$ and $\tau_c\to z^{-1/2}$; they trivially hold for $c=+\infty$. Then $|\Pi_c|^2 = |\tau_c|+c^{-2}|\tau_c|^{-1}\le M_0$ for $c\ge 1$, $|\Delta_c|^2 = |\tau_c|+|\tau_c|^{-1}\le M_0$, and $|Q_c|\le|\Delta_c|^2|\Ah| = |\Delta_c|^2|A|$.

(b) This follows from $|I\pm\sigma_1| = 2$ and $|\ee^{i\kappa_c|x|}| = \ee^{-\Im\kappa_c|x|}$.

(c) The first bound in the minimum follows from (b). For the second one, write
\[
F_c(x-h)-F_c(x) = \big(I+\sgn(x-h)\sigma_1\big)\big(\ee^{i\kappa_c|x-h|}-\ee^{i\kappa_c|x|}\big) + \big(\sgn(x-h)-\sgn(x)\big)\sigma_1\ee^{i\kappa_c|x|}.
\]
For $a,b\ge0$ we have $|\ee^{i\kappa_c a}-\ee^{i\kappa_c b}| = |\kappa_c\int_b^a\ee^{i\kappa_ct}\dd t|\le|\kappa_c||a-b|\ee^{-\gamma\min\{a,b\}}$. Since $||x-h|-|x||\le|h|$ and $\min\{|x-h|,|x|\}\ge|x|-|h|$, the first term is bounded by $2K|h|\ee^{-\gamma(|x|-|h|)}$. The second term vanishes unless $x$ lies between $0$ and $h$, and it is bounded by $2\sqrt2$. Hence
\[
\|\tau_hF_c-F_c\|_2^2\le 8K^2h^2\int_\R\ee^{-2\gamma(|x|-|h|)}\dd x + 16|h| = 8K^2h^2\gamma^{-1}\ee^{2\gamma|h|}+16|h|.
\]
Since $\|\tau_h\check F_c-\check F_c\|_2 = \|\tau_{-h}F_c-F_c\|_2$ and $\omega$ is even, the claim for $\check F_c$ follows.

(d) Since $\int v=1$, $\alpha_1(\epsilon\kappa_c)-1 = \int_{\R^2}v(x)v(y)(\ee^{i\epsilon\kappa_c|x-y|}-1)\dd x\dd y$, and $|\ee^{iu}-1|\le\min\{2,|u|\}$ for $\Im u\ge0$. The convergence $\eta(\epsilon)\to0$ follows from the dominated convergence theorem.

(e) By (b), $|F_c(x)-F_\infty(x)|^2\le16\ee^{-2\gamma|x|}$, and $F_c(x)\to F_\infty(x)$ pointwise. The dominated convergence theorem applies.

(f) For $|h|\le1$ we have $h^2\ee^{2\gamma|h|}\le\ee^{2\gamma}|h|$, and for $|h|>1$ we have $4\gamma^{-1/2}\le4\gamma^{-1/2}|h|^{1/2}$. Hence the first claim holds with $C_\omega := \max\{(8K^2\gamma^{-1}\ee^{2\gamma}+16)^{1/2},4\gamma^{-1/2}\}$. The second claim follows from $\min\{2,a\}\le(2a)^{1/2}$ for $a\ge0$ and $|x-y|^{1/2}\le|x|^{1/2}+|y|^{1/2}$.

(g) Let $s_c := (1+z/c^2)^{1/2}$ be the principal square root. For $c$ large, $\sqrt z\,s_c$ has a positive imaginary part and $(\sqrt z\,s_c)^2 = \lambda_c$; hence $\kappa_c = \sqrt z\,s_c$ and $\tau_c = s_c^2/\kappa_c = s_c/\sqrt z$. Since $|s_c-1| = |s_c^2-1|/|s_c+1|\le|z|c^{-2}$, we get $|\kappa_c-\sqrt z|+|\tau_c-z^{-1/2}| = O(c^{-2})$. As $\Re\vartheta_\infty>0$, the principal square root and its reciprocal are Lipschitz near $\tau_\infty = z^{-1/2}$, hence $|\Delta_c-\Delta_\infty| = O(c^{-2})$, which gives $|Q_c-Q_\infty| = O(c^{-2})$, and $\Pi_c-\Pi_\infty = \diag(\vartheta_c-\vartheta_\infty,c^{-1}\vartheta_c^{-1}) = O(c^{-1})$. Next, $\kappa_c$ and $\sqrt z$ lie in the convex set $\{w\mid\Im w\ge\gamma\}$, so $|\ee^{i\kappa_c|x|}-\ee^{i\sqrt z|x|}|\le|\kappa_c-\sqrt z||x|\ee^{-\gamma|x|}$, which yields the bounds on $F_c-F_\infty$. Finally, $\alpha_1$ is analytic in the open upper half-plane with $\alpha_1'(w) = i\int_{\R^2}|x-y|v(x)v(y)\ee^{iw|x-y|}\dd x\dd y$; since $r\ee^{-\delta r}\le(\ee\delta)^{-1}$ for $r\ge0$, we have $|\alpha_1'(w)|\le\|v\|_1^2/(\ee\delta)$ whenever $\Im w\ge\delta>0$. Integrating along the segment between $\epsilon\kappa_c$ and $\epsilon\sqrt z$, which lies in $\{\Im w\ge\epsilon\gamma\}$, we get the last estimate. Enlarging $c_0$ if necessary, all these bounds hold for $c\ge c_0$.
\end{proof}

\section{Uniform convergence of the non-local approximations}\label{sec:uniform}

We are now in position to prove our first main result.

\begin{theorem}\label{thm:uniform}
Let $v$ satisfy Assumption~\ref{ass:v}, $A\in\C^{2,2}$, and let $z\in\C\setminus[0,+\infty)$ be such that $\det M_z\neq 0$, i.e., $z\in\res(H_A)$. Then there are $c_1>0$ and $\epsilon_1>0$ such that for all $c\in[c_1,+\infty)$ and $\epsilon\in(0,\epsilon_1)$,
\[
z+\frac{c^2}{2}\in\res(D_{A,c})\cap\res(D_{A,c}^\epsilon),
\]
the operator $(D^\epsilon_{A,c}-c^2/2-z)^{-1} - (D_{A,c}-c^2/2-z)^{-1}$ is Hilbert--Schmidt, and
\begin{equation}\label{eq:uniform-estimate}
\sup_{c\ge c_1}\Big\|\Big(D^\epsilon_{A,c}-\frac{c^2}{2}-z\Big)^{-1}-\Big(D_{A,c}-\frac{c^2}{2}-z\Big)^{-1}\Big\|_{\HS}\le C\big(\theta(\epsilon)+\eta(\epsilon)\big)\xrightarrow{\epsilon\to0+}0,
\end{equation}
where $C>0$ is independent of $\epsilon$ and $c$, $\eta$ is as in Lemma~\ref{lem:uniform-bounds} and
\[
\theta(\epsilon) := \int_\R|v(s)|\,\omega(\epsilon s)\dd s.
\]
In particular,
\[
\lim_{\epsilon\to0+}\sup_{c\ge c_1}\Big\|\Big(D^\epsilon_{A,c}-\frac{c^2}{2}-z\Big)^{-1}-\Big(D_{A,c}-\frac{c^2}{2}-z\Big)^{-1}\Big\| = 0.
\]
Moreover, $d_\infty(\alpha_1(\epsilon\sqrt z)) = \det\big(I+\frac{i}{2}\alpha_1(\epsilon\sqrt{z})A\Zt(z)\big)\neq0$ for all $\epsilon\in(0,\epsilon_1)$. Finally, if $m_v<\infty$, then
\begin{equation}\label{eq:rate-eps}
\theta(\epsilon)+\eta(\epsilon)\le\big(C_\omega+2(2K)^{1/2}\|v\|_1\big)m_v\,\epsilon^{1/2},\qquad\epsilon>0,
\end{equation}
i.e., the convergence holds with the rate $O(\epsilon^{1/2})$, uniformly in $c\ge c_1$.
\end{theorem}

\begin{proof}
Let $c_0$ be as in Lemma~\ref{lem:uniform-bounds} and let $c\in[c_0,+\infty]$. Throughout the proof, $C_1,C_2,\dots$ denote positive constants which depend on $z$, $A$ and $v$, but not on $c$ and $\epsilon$.

By \eqref{eq:dc}, $d_c(\alpha) = 1+\frac{i}{2}\alpha t_c - \frac{\alpha^2}{4}\det A$ with $t_c := a_{11}\tau_c+a_{22}\tau_c^{-1}$, and $|t_c|\le|A|M_0$ by the Cauchy--Schwarz inequality and Lemma~\ref{lem:uniform-bounds}(a). Put $\varrho := \max\{2,\|v\|_1^2\}$. For $|\alpha|\le\varrho$ we have
\[
|d_c(\alpha)-d_c(1)| = |\alpha-1|\,\Big|\frac{i}{2}t_c-\frac{\alpha+1}{4}\det A\Big|\le C_1|\alpha-1|.
\]
Since $d_c(1)\to d_\infty(1) = \det M_z\neq0$ as $c\to+\infty$, there is $c_1\ge c_0$ such that $|d_c(1)|\ge\frac34|\det M_z|$ for all $c\in[c_1,+\infty]$. Choose $\epsilon_1>0$ such that $C_1\eta(\epsilon)\le\frac14|\det M_z|$ for $\epsilon\in(0,\epsilon_1)$. Since $|\alpha_1(\epsilon\kappa_c)|\le\|v\|_1^2\le\varrho$, Lemma~\ref{lem:uniform-bounds}(d) yields
\begin{equation}\label{eq:det-lower-bound}
|d_c(1)|\ge\frac12|\det M_z|\quad\text{and}\quad|d_c(\alpha_1(\epsilon\kappa_c))|\ge\frac12|\det M_z|,\qquad c\in[c_1,+\infty],\ \epsilon\in(0,\epsilon_1).
\end{equation}
For $c<+\infty$, Lemma~\ref{lem:factorization} implies that $w_c = z+c^2/2$ belongs to $\res(D_{A,c})\cap\res(D^\epsilon_{A,c})$. For $c=+\infty$, \eqref{eq:det-lower-bound} is the last assertion of the theorem.

In what follows, let $c\in[c_1,+\infty)$ and $\epsilon\in(0,\epsilon_1)$, and abbreviate $\alpha_\epsilon := \alpha_1(\epsilon\kappa_c)$, $N_\epsilon:=N_c(\alpha_\epsilon)$ and $N:=N_c(1)$. By \eqref{eq:inverse-2x2}, \eqref{eq:det-lower-bound} and Lemma~\ref{lem:uniform-bounds}(a), for $\alpha\in\{1,\alpha_\epsilon\}$,
\[
\Big|\Big(I+\frac{i}{2}\alpha Q_c\Big)^{-1}\Big|\le\frac{\sqrt2+\frac{\varrho}{2}M_0|A|}{\frac12|\det M_z|} =: C_2,\qquad |N_c(\alpha)|\le C_2M_0|A| =: C_3.
\]
Moreover, the identity
\begin{align*}
N_\epsilon - N &= \Big(I+\frac{i}{2}\alpha_\epsilon Q_c\Big)^{-1}\Big[\Big(I+\frac{i}{2}Q_c\Big)-\Big(I+\frac{i}{2}\alpha_\epsilon Q_c\Big)\Big]\Big(I+\frac{i}{2}Q_c\Big)^{-1}Q_c\\
&= \frac{i}{2}(1-\alpha_\epsilon)\Big(I+\frac{i}{2}\alpha_\epsilon Q_c\Big)^{-1}Q_c\,N
\end{align*}
together with Lemma~\ref{lem:uniform-bounds}(d) yields
\begin{equation}\label{eq:N-difference}
|N_\epsilon-N|\le\frac12C_2M_0|A|C_3\,\eta(\epsilon) =: C_4\,\eta(\epsilon).
\end{equation}

By Lemma~\ref{lem:factorization}, the difference of the resolvents is the integral operator with the kernel
\[
\mathcal K^c_\epsilon(x,y) = \frac14\Pi_c\Big[(v_\epsilon*F_c)(x)N_\epsilon(v_\epsilon*\check F_c)(y)-F_c(x)N\check F_c(y)\Big]\Pi_c.
\]
We decompose
\begin{multline*}
(v_\epsilon*F_c)(x)N_\epsilon(v_\epsilon*\check F_c)(y)-F_c(x)N\check F_c(y) = \big(v_\epsilon*F_c-F_c\big)(x)\,N_\epsilon\,(v_\epsilon*\check F_c)(y)\\
+F_c(x)\,(N_\epsilon-N)\,(v_\epsilon*\check F_c)(y) + F_c(x)\,N\,\big(v_\epsilon*\check F_c-\check F_c\big)(y).
\end{multline*}
By Lemma~\ref{lem:elementary}(c), $\int v_\epsilon = 1$ and Lemma~\ref{lem:uniform-bounds}(c),
\[
\|v_\epsilon*F_c-F_c\|_2\le\int_\R|v_\epsilon(s)|\,\omega(s)\dd s = \theta(\epsilon),\qquad\|v_\epsilon*\check F_c-\check F_c\|_2\le\theta(\epsilon),
\]
and $\|v_\epsilon*\check F_c\|_2\le\|v\|_1\|\check F_c\|_2\le2\|v\|_1\gamma^{-1/2}$. Using Lemma~\ref{lem:elementary}(a), $|\Pi_c|^2\le M_0$, Lemma~\ref{lem:uniform-bounds}(b) and \eqref{eq:N-difference}, we arrive at
\[
\|\mathcal K^c_\epsilon\|_\HS\le\frac{M_0}{4}\Big(2C_3\|v\|_1\gamma^{-1/2}\theta(\epsilon)+4\|v\|_1\gamma^{-1}C_4\eta(\epsilon)+2\gamma^{-1/2}C_3\theta(\epsilon)\Big)\le C\big(\theta(\epsilon)+\eta(\epsilon)\big).
\]
This bound does not depend on $c\in[c_1,+\infty)$. Finally, $\omega(\epsilon s)\to0$ as $\epsilon\to0+$ for every $s\in\R$ and $\omega\le4\gamma^{-1/2}$, so $\theta(\epsilon)\to0$ by the dominated convergence theorem. Since the operator norm does not exceed the Hilbert--Schmidt norm, the convergence follows. Finally, if $m_v<\infty$, then Lemma~\ref{lem:uniform-bounds}(f) gives $\theta(\epsilon)\le C_\omega\epsilon^{1/2}\int_\R|s|^{1/2}|v(s)|\dd s$ and the bound on $\eta(\epsilon)$, which yields \eqref{eq:rate-eps}.
\end{proof}

\begin{remark}
If $\int_\R|s||v(s)|\dd s<\infty$, then $\eta(\epsilon) = O(\epsilon)$, but $\theta(\epsilon)$ is in general only $O(\epsilon^{1/2})$. This is caused by the jump of $F_c$ at the origin, i.e., by the jump of the free resolvent kernel on the diagonal; for this reason we expect the exponent $1/2$ to be optimal for the Hilbert--Schmidt norm. We do not know whether it is optimal also for the operator norm.
\end{remark}

\begin{remark}\label{rem:c-infty}
The proof of Theorem~\ref{thm:uniform} works verbatim also for $c=+\infty$, i.e., the operator with the kernel
\[
\frac14\Pi_\infty\Big[(v_\epsilon*F_\infty)(x)N_\infty(\alpha_1(\epsilon\sqrt z))(v_\epsilon*\check F_\infty)(y)-F_\infty(x)N_\infty(1)\check F_\infty(y)\Big]\Pi_\infty
\]
satisfies the same Hilbert--Schmidt estimate \eqref{eq:uniform-estimate}. By Lemma~\ref{lem:c-infinity} and Theorem~\ref{thm:identification} below, this is exactly $P_1\otimes\big((H^\epsilon_A-z)^{-1}-(H_A-z)^{-1}\big)$.
\end{remark}

\begin{remark}
For a fixed $c$, Theorem~\ref{thm:uniform} reproduces \eqref{eq:rel-nonlocal-limit} at the point $w_c$. Note, however, that the uniformity cannot be obtained from the estimates of \cite{HeTu22} directly: the free resolvent kernel $R_c(w_c;\cdot,\cdot)$ contains the matrix $Z_c(w_c)$ whose norm grows linearly in $c$, and the same is true for the coupling matrix $A_c$. The factorization of Lemma~\ref{lem:factorization} redistributes these growing factors in such a way that only the bounded matrices $\Pi_c$, $Q_c$ and $N_c$ remain.
\end{remark}

\section{Non-relativistic limit of the non-local approximations}\label{sec:nonrel}

\subsection{The limit for a fixed $\epsilon$}
First we recall the non-relativistic limit of the free Dirac operator, which is classical, see \cite{GGT84} and \cite[Cor.~6.2]{Th92}. With the factorization of Lemma~\ref{lem:factorization} at hand, its proof, including the optimal rate $O(c^{-1})$, is very short.

\begin{lemma}\label{lem:free-nonrel}
Let $z\in\C\setminus[0,+\infty)$. Then
\[
\Big\|\Big(D_{0,c}-\frac{c^2}{2}-z\Big)^{-1}-P_1\otimes(H_0-z)^{-1}\Big\| \le Cc^{-1},\qquad c\ge c_0,
\]
where $C>0$ does not depend on $c$.
\end{lemma}

\begin{proof}
By Lemma~\ref{lem:factorization}(a) and Lemma~\ref{lem:c-infinity}(a), the difference is the convolution operator with the kernel $k_c(x-y)$, where $k_c(x) := \frac{i}{2}\big(\Pi_cF_c(x)\Pi_c - \Pi_\infty F_\infty(x)\Pi_\infty\big)$. We have
\[
\Pi_cF_c\Pi_c-\Pi_\infty F_\infty\Pi_\infty = (\Pi_c-\Pi_\infty)F_c\Pi_c+\Pi_\infty F_c(\Pi_c-\Pi_\infty)+\Pi_\infty(F_c-F_\infty)\Pi_\infty,
\]
hence $\|k_c\|_1\le\frac12\big(2M_0^{1/2}|\Pi_c-\Pi_\infty|\,\|F_c\|_1+M_0\|F_c-F_\infty\|_1\big)$ with $\|F_c\|_1\le4\gamma^{-1}$. By Lemma~\ref{lem:uniform-bounds}(g), $\|k_c\|_1 = O(c^{-1})$ and the assertion follows from Lemma~\ref{lem:elementary}(b).
\end{proof}

For $\epsilon>0$ and $z\in\C\setminus[0,+\infty)$ we put
\begin{equation}\label{eq:Meps}
\begin{gathered}
M_\epsilon(z) := I+A\Theta_\epsilon(z),\qquad \Theta_\epsilon(z):=\frac{i}{2}\alpha_1(\epsilon\sqrt{z})\Zt(z),\\
\varphi_\epsilon := K_0(z)v_\epsilon\in H^2(\R),\qquad \Phi_\epsilon:=(\varphi_\epsilon,\varphi_\epsilon').
\end{gathered}
\end{equation}
Note that $\det M_\epsilon(z) = d_\infty(\alpha_1(\epsilon\sqrt z))$. If $\det M_\epsilon(z)\neq0$, we define the bounded operator $L_\epsilon(z)$ on $L^2(\R)$ by its kernel
\begin{equation}\label{eq:Leps}
L_\epsilon(z;x,y) := K_0(z;x,y) - \Phi_\epsilon(x)\,M_\epsilon(z)^{-1}A\,\Phi_\epsilon(y)^T.
\end{equation}
Since $\varphi_\epsilon,\varphi_\epsilon'\in L^2(\R)$, $L_\epsilon(z)$ is a rank at most two perturbation of $K_0(z)$.

\begin{theorem}\label{thm:nonrel-eps}
Let $v$ satisfy Assumption~\ref{ass:v}, $A\in\C^{2,2}$, $z\in\C\setminus[0,+\infty)$, and let $\epsilon>0$ be such that $\det M_\epsilon(z)\neq0$. Then $z+c^2/2\in\res(D^\epsilon_{A,c})$ for all sufficiently large $c$, and
\[
\lim_{c\to+\infty}\Big\|\Big(D^\epsilon_{A,c}-\frac{c^2}{2}-z\Big)^{-1}-P_1\otimes L_\epsilon(z)\Big\| = 0.
\]
More precisely, there are $C_\epsilon>0$ and $c_\epsilon>0$ such that the norm above does not exceed $C_\epsilon c^{-1}$ for all $c\ge c_\epsilon$.
\end{theorem}

\begin{proof}
By Lemma~\ref{lem:asymptotics} and the continuity of $\alpha_1$, $\alpha_1(\epsilon\kappa_c)\to\alpha_1(\epsilon\sqrt z)$, hence $d_c(\alpha_1(\epsilon\kappa_c))\to d_\infty(\alpha_1(\epsilon\sqrt z)) = \det M_\epsilon(z)\neq0$ and $N_c(\alpha_1(\epsilon\kappa_c))\to N_\infty(\alpha_1(\epsilon\sqrt z))$ as $c\to+\infty$. In particular, by Lemma~\ref{lem:factorization}(c), $w_c\in\res(D^\epsilon_{A,c})$ for $c$ large. Moreover, by Lemma~\ref{lem:elementary}(c) and Lemma~\ref{lem:uniform-bounds}(e),
\[
\|v_\epsilon*F_c-v_\epsilon*F_\infty\|_2\le\|v\|_1\|F_c-F_\infty\|_2\to0,\qquad \|v_\epsilon*\check F_c-v_\epsilon*\check F_\infty\|_2\to0,
\]
and $\Pi_c\to\Pi_\infty$. Writing the difference of the products as a telescopic sum and using Lemma~\ref{lem:elementary}(a), we infer that the integral operator with the kernel
\[
\frac14\Pi_c(v_\epsilon*F_c)(x)N_c(\alpha_1(\epsilon\kappa_c))(v_\epsilon*\check F_c)(y)\Pi_c
\]
converges in the Hilbert--Schmidt norm to the one with the kernel
\begin{equation}\label{eq:limit-kernel}
\frac14\Pi_\infty(v_\epsilon*F_\infty)(x)N_\infty(\alpha_1(\epsilon\sqrt z))(v_\epsilon*\check F_\infty)(y)\Pi_\infty.
\end{equation}
Together with Lemma~\ref{lem:factorization}(c) and Lemma~\ref{lem:free-nonrel}, this proves the norm convergence to $P_1\otimes K_0(z)$ plus the operator with kernel \eqref{eq:limit-kernel}.

To obtain the rate, note that $|Q_c-Q_\infty| = O(c^{-2})$ and $|\alpha_1(\epsilon\kappa_c)-\alpha_1(\epsilon\sqrt z)| = O(c^{-2})$ by Lemma~\ref{lem:uniform-bounds}(g). Since $d_c(\alpha_1(\epsilon\kappa_c))$ is bounded away from zero for large $c$, \eqref{eq:inverse-2x2} and the second resolvent identity for matrices yield $|N_c(\alpha_1(\epsilon\kappa_c))-N_\infty(\alpha_1(\epsilon\sqrt z))| = O(c^{-2})$. Moreover, $|\Pi_c-\Pi_\infty| = O(c^{-1})$ and $\|v_\epsilon*F_c-v_\epsilon*F_\infty\|_2\le\|v\|_1\|F_c-F_\infty\|_2 = O(c^{-2})$, and similarly for $\check F_c$. Hence the telescopic decomposition together with Lemma~\ref{lem:elementary}(a) shows that the Hilbert--Schmidt norm of the difference of the singular parts is $O(c^{-1})$. Combined with Lemma~\ref{lem:free-nonrel}, this gives the asserted rate.

It remains to compute \eqref{eq:limit-kernel}. Put $e(x):=\ee^{i\sqrt z|x|}$ and $\varsigma(x):=\sgn(x)\ee^{i\sqrt z|x|}$. As in the proof of Lemma~\ref{lem:c-infinity}, only the $(1,1)$-entry of \eqref{eq:limit-kernel} is non-zero, the first row of $(v_\epsilon*F_\infty)(x)$ is $(a_\epsilon(x),b_\epsilon(x))$ with $a_\epsilon:=v_\epsilon*e$ and $b_\epsilon:=v_\epsilon*\varsigma$, and, since $e$ is even and $\varsigma$ is odd, the first column of $(v_\epsilon*\check F_\infty)(y)$ is $(a_\epsilon(y),-b_\epsilon(y))^T$. By Lemma~\ref{lem:c-infinity}(b) we get, exactly as in the proof of Lemma~\ref{lem:c-infinity}(c),
\[
\eqref{eq:limit-kernel} = \frac14\big(z^{-1/2}a_\epsilon(x),\,ib_\epsilon(x)\big)M_\epsilon(z)^{-1}A\begin{pmatrix}z^{-1/2}a_\epsilon(y)\\ ib_\epsilon(y)\end{pmatrix}P_1.
\]
Finally, by \eqref{eq:free-schr-res}, $\varphi_\epsilon = \frac{i}{2\sqrt z}a_\epsilon$, and differentiating under the integral sign, $\varphi_\epsilon' = \frac{i}{2\sqrt z}\,i\sqrt z\,b_\epsilon = -\frac12b_\epsilon$. Hence $z^{-1/2}a_\epsilon = -2i\varphi_\epsilon$ and $ib_\epsilon = -2i\varphi_\epsilon'$, and \eqref{eq:limit-kernel} equals $-\Phi_\epsilon(x)M_\epsilon(z)^{-1}A\Phi_\epsilon(y)^TP_1$.
\end{proof}

\begin{remark}\label{rem:rate-point}
Replacing $v_\epsilon*F_c$, $v_\epsilon*\check F_c$ and $\alpha_1(\epsilon\kappa_c)$ by $F_c$, $\check F_c$ and $1$, respectively, and using Lemma~\ref{lem:factorization}(b) and Lemma~\ref{lem:c-infinity}(c) instead of Lemma~\ref{lem:factorization}(c), the same argument shows that, for $z\in\res(H_A)\setminus[0,+\infty)$,
\[
\Big\|\Big(D_{A,c}-\frac{c^2}{2}-z\Big)^{-1}-P_1\otimes(H_A-z)^{-1}\Big\|\le Cc^{-1}
\]
for all sufficiently large $c$, i.e., Theorem~\ref{thm:HeTu-nonrel} holds with the rate $O(c^{-1})$. This order cannot be improved in general: already for $A=0$, the off-diagonal entries of the kernel $R_c(w_c;x,y) = \frac{i}{2}\Pi_cF_c(x-y)\Pi_c$ equal $\frac{i}{2c}\sgn(x-y)\ee^{i\kappa_c|x-y|}$.
\end{remark}

\subsection*{The limit as a pseudoresolvent}
Theorem~\ref{thm:nonrel-eps} identifies the non-relativistic limit of the resolvents of $D^\epsilon_{A,c}$, but it does not tell us yet whether $L_\epsilon(z)$ is the resolvent of an operator. This question can be answered abstractly. Recall that a family $\{J(z)\}_{z\in\Omega}$ of bounded operators on a Banach space is called a \emph{pseudoresolvent} if it satisfies the resolvent identity $J(z)-J(w) = (z-w)J(z)J(w)$ for all $z,w\in\Omega$. Norm (and strong) limits of resolvents are always pseudoresolvents, and the question whether such a limit is again a resolvent is classical in approximation theorems of Trotter--Kato type, see, e.g., \cite[Ch.~VIII]{Ka95} and \cite[Ch.~III]{EnNa00}. In general, a pseudoresolvent $J$ is the resolvent of the closed linear relation (multivalued operator)
\[
T = \{(J(z)g,\,g+zJ(z)g)\mid g\in\mathcal X\},
\]
which does not depend on $z\in\Omega$ and whose multivalued part equals $\Ker J(z)$, see \cite{Ar61} and \cite[Ch.~1]{BHdS20}. Consequently, $J$ is the resolvent of an operator if and only if $\Ker J(z)=\{0\}$. For the reader's convenience, we include \cite[Lem. 4.5 \& Prop. 4.6]{EnNa00} as the following lemma with elementary proof.

\begin{lemma}\label{lem:pseudoresolvent}
Let $\mathcal X$ be a Banach space, $\emptyset\neq\Omega\subset\C$ and let
$J:\Omega\to\mathcal B(\mathcal X)$ satisfy
$J(z)-J(w)=(z-w)J(z)J(w)$ for all $z,w\in\Omega$. Then $\Ker J(z)$ and
$\Ran J(z)$ do not depend on $z\in\Omega$. If $\Ker J(z_0)=\{0\}$ for some
$z_0\in\Omega$, then $T:=z_0+J(z_0)^{-1}$, $\Dom T:=\Ran J(z_0)$, is a closed
operator independent of $z_0$, $\Omega\subset\res(T)$ and $(T-w)^{-1}=J(w)$
for all $w\in\Omega$. Moreover, $T$ is densely defined if and only if
$\Ran J(z_0)$ is dense.
\end{lemma}

\begin{proof}
If $J(z)g=0$, then $J(w)g=J(z)g+(w-z)J(w)J(z)g=0$. Since
$J(z)=J(w)\big(I+(z-w)J(z)\big)$, we have $\Ran J(z)\subset\Ran J(w)$, and by
symmetry the ranges coincide. Assume $\Ker J(z_0)=\{0\}$. Then $T$ is closed
as the inverse of a bounded injective operator, shifted by $z_0$. Let
$w\in\Omega$. From $J(w)=J(z_0)\big(I+(w-z_0)J(w)\big)$ we get
$J(z_0)^{-1}J(w)g=g+(w-z_0)J(w)g$, hence $(T-w)J(w)g=g$ for all
$g\in\mathcal X$. Conversely, for $\psi=J(z_0)h\in\Dom T$,
\[
J(w)(T-w)\psi=J(w)h+(z_0-w)J(w)J(z_0)h=J(w)h-\big(J(w)-J(z_0)\big)h=\psi .
\]
Thus $T-w$ is bijective with $(T-w)^{-1}=J(w)$. In particular, $T$ does not
depend on the choice of $z_0$.
\end{proof}

For $\epsilon>0$ we put $\Omega_\epsilon:=\{z\in\C\setminus[0,+\infty)\mid\det M_\epsilon(z)\neq0\}$, and we denote by $M^*_\epsilon$, $L^*_\epsilon$ and $\Omega^*_\epsilon$ the objects defined by \eqref{eq:Meps}, \eqref{eq:Leps} and above with $A$ replaced by $A^*$. For $g\in L^2(\R)$ we also put $p_\epsilon(g) := \int_\R\Phi_\epsilon(y)^Tg(y)\dd y\in\C^2$, so that $L_\epsilon(z)g = K_0(z)g-\Phi_\epsilon M_\epsilon(z)^{-1}Ap_\epsilon(g)$.

\begin{prop}\label{prop:Teps}
Let $v$ satisfy Assumption~\ref{ass:v} and let $\epsilon>0$ be such that $\Omega_\epsilon\neq\emptyset$.
\begin{enumerate}[(a)]
\item $\{L_\epsilon(z)\}_{z\in\Omega_\epsilon}$ is a pseudoresolvent on $L^2(\R)$.
\item For every $z\in\Omega_\epsilon$ we have $\bar z\in\Omega_\epsilon^*$ and $L_\epsilon(z)^* = L^*_\epsilon(\bar z)$.
\item If $v\notin H^1(\R)$, then $\Ker L_\epsilon(z)=\{0\}$ for all $z\in\Omega_\epsilon$, and there is a unique closed, densely defined operator $T_\epsilon$ in $L^2(\R)$ such that $\Omega_\epsilon\subset\res(T_\epsilon)$ and $(T_\epsilon-z)^{-1}=L_\epsilon(z)$ for all $z\in\Omega_\epsilon$. In particular, $D^\epsilon_{A,c}-c^2/2$ converges to $P_1\otimes T_\epsilon$ in the norm resolvent sense as $c\to+\infty$.
\end{enumerate}
\end{prop}

\begin{proof}
(a) Let $z,w\in\Omega_\epsilon$. By Theorem~\ref{thm:nonrel-eps}, for all sufficiently large $c$, both $z+c^2/2$ and $w+c^2/2$ belong to $\res(D^\epsilon_{A,c})$, and the resolvents of $D^\epsilon_{A,c}-c^2/2$ satisfy the resolvent identity. Passing to the norm limit as $c\to+\infty$ yields the identity for $P_1\otimes L_\epsilon=L_\epsilon\oplus0$, hence for $L_\epsilon$.

(b) Recall that $\sqrt{\bar z} = -\overline{\sqrt z}$. Consequently, $K_0(z;y,x) = K_0(z;x,y) = \overline{K_0(\bar z;x,y)}$, and, since $v$ is real, $\overline{\varphi_\epsilon(z)} = \varphi_\epsilon(\bar z)$, where we temporarily indicate the dependence on $z$. Moreover, $\alpha_1(\epsilon\sqrt{\bar z}) = \overline{\alpha_1(\epsilon\sqrt z)}$ and $\Zt(\bar z) = -\overline{\Zt(z)}$, whence $\Theta_\epsilon(\bar z) = \overline{\Theta_\epsilon(z)} = \Theta_\epsilon(z)^*$. Therefore $\det M^*_\epsilon(\bar z) = \det(I+A^*\Theta_\epsilon(z)^*) = \overline{\det M_\epsilon(z)}\neq0$, and, by the push-through identity, $(M_\epsilon(z)^{-1}A)^* = A^*(I+\Theta_\epsilon(z)^*A^*)^{-1} = M^*_\epsilon(\bar z)^{-1}A^*$. Taking the complex conjugate of the kernel \eqref{eq:Leps} with interchanged variables we arrive at $L_\epsilon(z)^* = L_\epsilon^*(\bar z)$.

(c) Let $z\in\Omega_\epsilon$ and $L_\epsilon(z)g=0$. Put $\eta := M_\epsilon(z)^{-1}Ap_\epsilon(g)$. Then $K_0(z)g = \Phi_\epsilon\eta = \eta_1\varphi_\epsilon+\eta_2\varphi_\epsilon'$. Since $(H_0-z)\varphi_\epsilon = v_\epsilon$, we have $(H_0-z)\varphi_\epsilon' = v_\epsilon'$ in the sense of distributions, and so $g = \eta_1v_\epsilon+\eta_2v_\epsilon'$. As $g,v_\epsilon\in L^2(\R)$ and $v_\epsilon'\notin L^2(\R)$, we get $\eta_2=0$. Furthermore, $\int_\R\varphi_\epsilon v_\epsilon = \frac{i}{2\sqrt z}\alpha_1(\epsilon\sqrt z) = \Theta_\epsilon(z)_{11}$ and $\int_\R\varphi_\epsilon'v_\epsilon = 0$, because $\varphi_\epsilon'(x) = -\frac12\int_\R\sgn(x-y)\ee^{i\sqrt z|x-y|}v_\epsilon(y)\dd y$ (cf.~the proof of Theorem~\ref{thm:nonrel-eps}) and the resulting double integral has an antisymmetric integrand. Hence $p_\epsilon(g) = \eta_1p_\epsilon(v_\epsilon) = \Theta_\epsilon(z)\eta$ and
\[
0 = \eta-M_\epsilon(z)^{-1}A\Theta_\epsilon(z)\eta = M_\epsilon(z)^{-1}\big(M_\epsilon(z)-A\Theta_\epsilon(z)\big)\eta = M_\epsilon(z)^{-1}\eta.
\]
Thus $\eta=0$ and $g=0$. By (b), the same argument applied to $A^*$ gives $\Ker L_\epsilon(z)^* = \Ker L^*_\epsilon(\bar z) = \{0\}$, i.e., $\Ran L_\epsilon(z)$ is dense. The assertion now follows from (a), Lemma~\ref{lem:pseudoresolvent} and Theorem~\ref{thm:nonrel-eps}.
\end{proof}

\begin{remark}\label{rem:Omega-eps}
\begin{enumerate}[(a)]
\item By Theorem~\ref{thm:uniform}, every $z\in\res(H_A)\setminus[0,+\infty)$ belongs to $\Omega_\epsilon$ for all sufficiently small $\epsilon$. Moreover, for $t>0$ we have $\sqrt{-t}=i\sqrt t$ and
\[
\Theta_\epsilon(-t) = \diag\Big(\frac{\alpha_1(i\epsilon\sqrt t)}{2\sqrt t},\,-\frac{\sqrt t}{2}\alpha_1(i\epsilon\sqrt t)\Big).
\] Since $\frac{s}{2}\ee^{-s|\cdot|}$ is an approximate identity, $\frac{s}{2}\alpha_1(is)\to\|v\|_2^2$ as $s\to+\infty$, whence
\[
\lim_{t\to+\infty}M_\epsilon(-t) = I-n_\epsilon AE_{22},\qquad \lim_{t\to+\infty}\det M_\epsilon(-t) = 1-a_{22}n_\epsilon.
\]
In particular, $\Omega_\epsilon\neq\emptyset$ whenever $1-a_{22}n_\epsilon\neq0$. The meaning of the number $1-a_{22}n_\epsilon$ will become clear in Proposition~\ref{prop:kernel-H1}.
\item If $v\notin H^1(\R)$, then $v_\epsilon'$ belongs only to $H^{-1}(\R)$, and $T_\epsilon$ may be viewed as a singular finite-rank perturbation of $H_0$ in the sense of \cite{AlKu00}, formally given by the right-hand side of \eqref{eq:HAeps} below. We do not pursue the description of $\Dom T_\epsilon$ here.
\end{enumerate}
\end{remark}

\subsection{Identification of the limit operator}

In order to identify $L_\epsilon(z)$ in the remaining case $v\in H^1(\R)$ as
the resolvent of a bounded rank-two perturbation of the free Schrödinger
operator, we strengthen Assumption~\ref{ass:v}.

\begin{assumption}\label{ass:v-prime}
In addition to Assumption~\ref{ass:v}, $v\in H^1(\R)$.
\end{assumption}

Under Assumption~\ref{ass:v-prime}, $v_\epsilon'\in L^2(\R;\R)$ and we put
\[
u_1^\epsilon := v_\epsilon,\qquad u_2^\epsilon := v_\epsilon',
\]
and recall that $n_\epsilon = \|v_\epsilon\|_2^2 = \epsilon^{-1}\|v\|_2^2$.
If $1-a_{22}n_\epsilon\neq0$, we define
\begin{equation}\label{eq:Beps}
B_\epsilon := \big(I-n_\epsilon AE_{22}\big)^{-1}A = \frac{1}{1-a_{22}n_\epsilon}\begin{pmatrix} a_{11}-n_\epsilon\det A & a_{12}\\ a_{21} & a_{22}\end{pmatrix}
\end{equation}
and
\begin{equation}\label{eq:HAeps}
\Dom H_A^\epsilon := H^2(\R),\qquad H^\epsilon_A\psi := -\psi''+\sum_{j,k=1}^2(B_\epsilon)_{jk}\langle u_k^\epsilon,\psi\rangle u_j^\epsilon.
\end{equation}
The operator $H_A^\epsilon$ is a bounded rank-two perturbation of $H_0$; in particular, it is closed. Since $a_{22}n_\epsilon\to\infty$ if $a_{22}\neq0$, the condition $1-a_{22}n_\epsilon\neq0$ holds for all sufficiently small $\epsilon>0$ (it can fail for at most one value of $\epsilon$).

\begin{remark}
Formally, $u_1^\epsilon\to\delta$ and $u_2^\epsilon\to\delta'$, so that $H_A^\epsilon$ has the same structure as the formal expression \eqref{eq:formal-H}, but with $A$ replaced by $B_\epsilon$. If $A$ is hermitian, then so is $B_\epsilon$ (note that $a_{22}$ and $\det A$ are real), and $H^\epsilon_A$ is self-adjoint. For the $\delta$-interaction, $A = aE_{11}$, we get $B_\epsilon = A$ and $H^\epsilon_A = H_0+a\ket{v_\epsilon}\bra{v_\epsilon}$, which is the well-known non-local approximation of the $\delta$-interaction, cf.~\cite{AGHH05}. For the $\delta'$-interaction, $A = bE_{22}$, we get
\[
H_A^\epsilon = H_0+\frac{b}{1-b\,\epsilon^{-1}\|v\|_2^2}\ket{v_\epsilon'}\bra{v_\epsilon'},
\]
i.e., the coupling constant has to be renormalized; this is in accordance with \cite{Se86b}, where non-local approximations of the $\delta'$-interaction of this type were studied.
\end{remark}

We will use the following finite-rank version of the Krein resolvent formula, which we prove for completeness.

\begin{lemma}\label{lem:finite-rank}
Let $u_1,u_2\in L^2(\R)$, $B\in\C^{2,2}$, and let $U:\C^2\to L^2(\R)$, $U\xi := \xi_1u_1+\xi_2u_2$. Let $z\in\res(H_0)$ and let $G\in\C^{2,2}$ be the matrix with entries $G_{jk} := \langle u_j,K_0(z)u_k\rangle$, i.e., $G = U^*K_0(z)U$. If $\det(I+BG)\neq0$, then $z\in\res(H_0+UBU^*)$ and
\[
(H_0+UBU^*-z)^{-1} = K_0(z)-K_0(z)U(I+BG)^{-1}BU^*K_0(z).
\]
\end{lemma}

\begin{proof}
On $\Dom H_0$ we have $H_0+UBU^*-z = (I+UBU^*K_0(z))(H_0-z)$. Put $X:=U^*K_0(z)$, so that $XU = G$. A direct computation shows
\[
(I+UBX)\big(I-U(I+BG)^{-1}BX\big) = I+UBX-U\big(I+BXU\big)(I+BG)^{-1}BX = I,
\]
and similarly $\big(I-U(I+BG)^{-1}BX\big)(I+UBX) = I$. Hence $I+UBX$ is a bijection of $L^2(\R)$ with a bounded inverse, which yields the claim.
\end{proof}

\begin{lemma}\label{lem:gram}
Let $v$ satisfy Assumption~\ref{ass:v-prime}, $\epsilon>0$ and $z\in\C\setminus[0,+\infty)$. Then $K_0(z)v_\epsilon' = \varphi_\epsilon'$ and
\[
G_\epsilon := \Big(\langle u_j^\epsilon,K_0(z)u_k^\epsilon\rangle\Big)_{j,k=1}^2 = \frac{i}{2}\alpha_1(\epsilon\sqrt z)\Zt(z)+n_\epsilon E_{22}.
\]
\end{lemma}

\begin{proof}
$K_0(z)$ acts as the convolution with an $L^1$-function, hence it commutes with differentiation on $H^1(\R)$ and $K_0(z)v_\epsilon' = (K_0(z)v_\epsilon)' = \varphi_\epsilon'$. Since $v_\epsilon$ is real,
\[
\langle v_\epsilon,K_0(z)v_\epsilon\rangle = \frac{i}{2\sqrt z}\int_{\R^2}v_\epsilon(x)\ee^{i\sqrt z|x-y|}v_\epsilon(y)\dd x\dd y = \frac{i}{2\sqrt z}\alpha_1(\epsilon\sqrt z).
\]
Next, by the proof of Theorem~\ref{thm:nonrel-eps}, $\varphi_\epsilon'(x) = -\frac12\int_\R\sgn(x-y)\ee^{i\sqrt z|x-y|}v_\epsilon(y)\dd y$, so $\langle v_\epsilon,K_0(z)v_\epsilon'\rangle = \langle v_\epsilon,\varphi_\epsilon'\rangle = 0$, because the integrand is antisymmetric with respect to the interchange of $x$ and $y$. Similarly, integrating by parts, $\langle v_\epsilon',K_0(z)v_\epsilon\rangle = -\langle v_\epsilon,\varphi_\epsilon'\rangle = 0$. Finally, since $\varphi_\epsilon\in H^2(\R)$ and $-\varphi_\epsilon'' = v_\epsilon + z\varphi_\epsilon$,
\[
\langle v_\epsilon',K_0(z)v_\epsilon'\rangle = \langle v_\epsilon',\varphi_\epsilon'\rangle = -\langle v_\epsilon,\varphi_\epsilon''\rangle = \|v_\epsilon\|_2^2+z\langle v_\epsilon,K_0(z)v_\epsilon\rangle = n_\epsilon+\frac{i\sqrt z}{2}\alpha_1(\epsilon\sqrt z).
\]
This is the asserted formula.
\end{proof}

\begin{theorem}\label{thm:identification}
Let $v$ satisfy Assumption~\ref{ass:v-prime}, $A\in\C^{2,2}$, $z\in\C\setminus[0,+\infty)$, and let $\epsilon>0$ be such that $1-a_{22}n_\epsilon\neq0$ and $\det M_\epsilon(z)\neq0$. Then $z\in\res(H_A^\epsilon)$ and $(H^\epsilon_A-z)^{-1} = L_\epsilon(z)$.
\end{theorem}

\begin{proof}
We have $H_A^\epsilon = H_0+UB_\epsilon U^*$ with $U$ as in Lemma~\ref{lem:finite-rank} for $u_j = u_j^\epsilon$. Put $\Theta := \Theta_\epsilon(z)$, so that $M_\epsilon(z) = I+A\Theta$ and $G_\epsilon = \Theta+n_\epsilon E_{22}$ by Lemma~\ref{lem:gram}. Then
\[
I+B_\epsilon G_\epsilon = (I-n_\epsilon AE_{22})^{-1}\big(I-n_\epsilon AE_{22}+A\Theta+n_\epsilon AE_{22}\big) = (I-n_\epsilon AE_{22})^{-1}M_\epsilon(z).
\]
Since $\det(I-n_\epsilon AE_{22}) = 1-a_{22}n_\epsilon\neq0$, we get $\det(I+B_\epsilon G_\epsilon) = \det M_\epsilon(z)/(1-a_{22}n_\epsilon)\neq0$ and
\[
(I+B_\epsilon G_\epsilon)^{-1}B_\epsilon = M_\epsilon(z)^{-1}(I-n_\epsilon AE_{22})(I-n_\epsilon AE_{22})^{-1}A = M_\epsilon(z)^{-1}A.
\]
By Lemma~\ref{lem:finite-rank}, $z\in\res(H^\epsilon_A)$ and $(H_A^\epsilon-z)^{-1} = K_0(z)-K_0(z)UM_\epsilon(z)^{-1}AU^*K_0(z)$. Since $K_0(z;x,y) = K_0(z;y,x)$ and $u_k^\epsilon$ is real, $\langle u_k^\epsilon,K_0(z)g\rangle = \int_\R(K_0(z)u_k^\epsilon)(y)g(y)\dd y$. Therefore, by Lemma~\ref{lem:gram}, the integral kernel of $K_0(z)UM_\epsilon(z)^{-1}AU^*K_0(z)$ is $\Phi_\epsilon(x)M_\epsilon(z)^{-1}A\Phi_\epsilon(y)^T$, which proves the theorem.
\end{proof}

Combining Theorems~\ref{thm:nonrel-eps} and~\ref{thm:identification} we obtain the non-relativistic limit of the non-local approximations.

\begin{cor}\label{cor:nonrel-eps}
Let $v$ satisfy Assumption~\ref{ass:v-prime}, $z\in\C\setminus[0,+\infty)$ and let $\epsilon>0$ be such that $1-a_{22}n_\epsilon\neq0$ and $\det M_\epsilon(z)\neq0$. Then
\[
\lim_{c\to+\infty}\Big\|\Big(D^\epsilon_{A,c}-\frac{c^2}{2}-z\Big)^{-1}-P_1\otimes(H^\epsilon_A-z)^{-1}\Big\| = 0,
\]
and the norm above is $O(c^{-1})$ as $c\to+\infty$.
\end{cor}

In the case $v\in H^1(\R)$, the kernel of the pseudoresolvent $L_\epsilon$ can be computed explicitly. It turns out that the exceptional value of $\epsilon$, for which $B_\epsilon$ is not defined, is exactly the value for which $L_\epsilon(z)$ is not the resolvent of an operator.

\begin{prop}\label{prop:kernel-H1}
Let $v$ satisfy Assumption~\ref{ass:v-prime}, $\epsilon>0$ and $z\in\Omega_\epsilon$. Let $U$ be as in Lemma~\ref{lem:finite-rank} with $u_j = u_j^\epsilon$. Then
\[
\Ker L_\epsilon(z) = U\Ker\big(I-n_\epsilon AE_{22}\big).
\]
Consequently:
\begin{enumerate}[(a)]
\item If $1-a_{22}n_\epsilon\neq0$, then $L_\epsilon(z) = (H^\epsilon_A-z)^{-1}$ for all $z\in\Omega_\epsilon$, by Theorem~\ref{thm:identification}.
\item If $1-a_{22}n_\epsilon=0$, then $\Ker L_\epsilon(z) = \operatorname{span}\{n_\epsilon a_{12}v_\epsilon+v_\epsilon'\}\neq\{0\}$. Hence $L_\epsilon$ is not the resolvent of any operator; it is the resolvent of a closed linear relation whose multivalued part is $\operatorname{span}\{n_\epsilon a_{12}v_\epsilon+v_\epsilon'\}$. In this case, the non-relativistic limit of $D^\epsilon_{A,c}$ from Theorem~\ref{thm:nonrel-eps} is not an operator.
\end{enumerate}
\end{prop}

\begin{proof}
By Lemma~\ref{lem:gram}, $K_0(z)U\eta = \Phi_\epsilon\eta$ and $p_\epsilon(U\eta) = G_\epsilon\eta$ for all $\eta\in\C^2$ (note that $G_\epsilon$ is symmetric). Let $L_\epsilon(z)g=0$ and put $\eta := M_\epsilon(z)^{-1}Ap_\epsilon(g)$. Then $K_0(z)g = \Phi_\epsilon\eta = K_0(z)U\eta$, i.e., $g = U\eta$, and $p_\epsilon(g) = G_\epsilon\eta$. Since $G_\epsilon = \Theta_\epsilon(z)+n_\epsilon E_{22}$, we get
\[
0 = \eta-M_\epsilon(z)^{-1}AG_\epsilon\eta = M_\epsilon(z)^{-1}\big(M_\epsilon(z)-A\Theta_\epsilon(z)-n_\epsilon AE_{22}\big)\eta = M_\epsilon(z)^{-1}\big(I-n_\epsilon AE_{22}\big)\eta.
\]
Conversely, for $\eta\in\Ker(I-n_\epsilon AE_{22})$ the same computation gives $L_\epsilon(z)U\eta = \Phi_\epsilon M_\epsilon(z)^{-1}(I-n_\epsilon AE_{22})\eta = 0$. This proves the formula for the kernel. If $1-a_{22}n_\epsilon=0$, then $I-n_\epsilon AE_{22} = \begin{pmatrix}1 & -n_\epsilon a_{12}\\ 0 & 0\end{pmatrix}$, whose kernel is spanned by $(n_\epsilon a_{12},1)^T$; since $v_\epsilon$ and $v_\epsilon'$ are linearly independent, $U$ is injective. The remaining assertions follow from the discussion preceding Lemma~\ref{lem:pseudoresolvent}.
\end{proof}

\section{Interchanging the limits}\label{sec:interchange}

Now we are ready to prove that the diagram \eqref{eq:diagram-intro} can be completed and that it commutes.

\begin{theorem}\label{thm:main}
Let $v$ satisfy Assumption~\ref{ass:v-prime}, $A\in\C^{2,2}$, and let $z\in\C\setminus[0,+\infty)$ be such that $z\in\res(H_A)$. Then there are $\epsilon_0>0$ and $c_1>0$ such that, for all $\epsilon\in(0,\epsilon_0)$ and $c\ge c_1$, we have $z+c^2/2\in\res(D_{A,c})\cap\res(D_{A,c}^\epsilon)$ and $z\in\res(H^\epsilon_A)$. Moreover, the following assertions hold.
\begin{enumerate}[(i)]
\item $\displaystyle\lim_{\epsilon\to0+}\sup_{c\ge c_1}\Big\|\Big(D^\epsilon_{A,c}-\frac{c^2}{2}-z\Big)^{-1}-\Big(D_{A,c}-\frac{c^2}{2}-z\Big)^{-1}\Big\| = 0.$
\item For every $\epsilon\in(0,\epsilon_0)$, $\displaystyle\lim_{c\to+\infty}\Big\|\Big(D^\epsilon_{A,c}-\frac{c^2}{2}-z\Big)^{-1}-P_1\otimes(H^\epsilon_A-z)^{-1}\Big\| = 0.$
\item $\displaystyle\lim_{c\to+\infty}\Big\|\Big(D_{A,c}-\frac{c^2}{2}-z\Big)^{-1}-P_1\otimes(H_A-z)^{-1}\Big\| = 0.$
\item $\displaystyle\lim_{\epsilon\to0+}\big\|(H^\epsilon_A-z)^{-1}-(H_A-z)^{-1}\big\| = 0.$
\item The double limit exists:
\[
\lim_{(\epsilon,c)\to(0+,+\infty)}\Big\|\Big(D^\epsilon_{A,c}-\frac{c^2}{2}-z\Big)^{-1}-P_1\otimes(H_A-z)^{-1}\Big\| = 0.
\]
\item (Rates.) The convergences in (ii) and (iii) are $O(c^{-1})$, and there is $C>0$ such that for all $\epsilon\in(0,\epsilon_0)$ and $c\ge c_1$,
\begin{gather*}
\big\|(H^\epsilon_A-z)^{-1}-(H_A-z)^{-1}\big\|\le C\big(\theta(\epsilon)+\eta(\epsilon)\big),\\
\Big\|\Big(D^\epsilon_{A,c}-\frac{c^2}{2}-z\Big)^{-1}-P_1\otimes(H_A-z)^{-1}\Big\|\le C\big(\theta(\epsilon)+\eta(\epsilon)+c^{-1}\big).
\end{gather*}
In particular, if $m_v<\infty$, these differences are $O(\epsilon^{1/2})$ and $O(\epsilon^{1/2}+c^{-1})$, respectively.
\end{enumerate}
In other words, the following diagram, in which all arrows denote the norm resolvent convergence, commutes:
\[
\begin{tikzcd}[column sep=6em, row sep=4em]
D_{A,c}^\epsilon - \frac{c^2}{2} \arrow[r, "\epsilon\to 0+", "\text{unif. in }c"'] \arrow[d, "c\to+\infty"'] & D_{A,c} - \frac{c^2}{2} \arrow[d, "c\to+\infty"] \\
H_A^\epsilon \arrow[r, "\epsilon\to 0+"'] & H_A
\end{tikzcd}
\]
\end{theorem}

\begin{proof}
Let $c_1$ and $\epsilon_1$ be as in Theorem~\ref{thm:uniform} and choose $\epsilon_0\in(0,\epsilon_1]$ such that $1-a_{22}n_\epsilon\neq0$ for $\epsilon\in(0,\epsilon_0)$. By Theorem~\ref{thm:uniform}, $\det M_\epsilon(z) = d_\infty(\alpha_1(\epsilon\sqrt z))\neq0$ for $\epsilon\in(0,\epsilon_0)$, so Theorem~\ref{thm:identification} gives $z\in\res(H^\epsilon_A)$. Assertion (i) is Theorem~\ref{thm:uniform}, (ii) is Corollary~\ref{cor:nonrel-eps}, and (iii) is Theorem~\ref{thm:HeTu-nonrel}.

Let $\mathcal X$ be the Banach space of bounded operators on $L^2(\R;\C^2)$ and put, for $\epsilon\in(0,\epsilon_0)$ and $c\ge c_1$,
\[
f(\epsilon,c):=\Big(D^\epsilon_{A,c}-\frac{c^2}{2}-z\Big)^{-1},\qquad g(c):=\Big(D_{A,c}-\frac{c^2}{2}-z\Big)^{-1},\qquad h(\epsilon):=P_1\otimes(H_A^\epsilon-z)^{-1}.
\]
By (i) and (ii), the assumptions of Lemma~\ref{lem:moore-osgood} are satisfied. Consequently, $\lim_{\epsilon\to0+}h(\epsilon) = \lim_{c\to+\infty}g(c)$, and the latter limit equals $P_1\otimes(H_A-z)^{-1}$ by (iii). Since $\|P_1\otimes T\| = \|T\|$ for every bounded operator $T$ on $L^2(\R)$, this is (iv). Assertion (v) is the last statement of Lemma~\ref{lem:moore-osgood}.

It remains to prove (vi). The rates in (ii) and (iii) are given by Corollary~\ref{cor:nonrel-eps} and Remark~\ref{rem:rate-point}; enlarging $c_1$ if necessary, $\|g(c)-P_1\otimes(H_A-z)^{-1}\|\le C_1c^{-1}$ for $c\ge c_1$. By (ii) and (iii),
\[
h(\epsilon)-P_1\otimes(H_A-z)^{-1} = \lim_{c\to+\infty}\big(f(\epsilon,c)-g(c)\big),
\]
so Theorem~\ref{thm:uniform} yields $\|(H^\epsilon_A-z)^{-1}-(H_A-z)^{-1}\|\le\sup_{c\ge c_1}\|f(\epsilon,c)-g(c)\|\le C(\theta(\epsilon)+\eta(\epsilon))$. In other words, the uniform rate of Theorem~\ref{thm:uniform} is inherited by the limit $c\to+\infty$. Similarly, $\|f(\epsilon,c)-P_1\otimes(H_A-z)^{-1}\|\le\|f(\epsilon,c)-g(c)\|+\|g(c)-P_1\otimes(H_A-z)^{-1}\|\le C(\theta(\epsilon)+\eta(\epsilon))+C_1c^{-1}$. The last claim follows from \eqref{eq:rate-eps}.
\end{proof}

\begin{remark}
\begin{enumerate}[(a)]
\item Assertion (v) means that the non-local approximations $D^\epsilon_{A,c}-c^2/2$ approximate the non-relativistic point interaction $H_A$ along \emph{any} path $(\epsilon,c)\to(0+,+\infty)$; e.g., one may take $\epsilon=\epsilon(c)$ with $\epsilon(c)\to0$ as $c\to+\infty$ arbitrarily fast or slowly. This is a genuinely stronger statement than the commutativity of the iterated limits and it relies on the uniformity in (i).
\item Assertions (i)--(iv) are not independent. By Remark~\ref{rem:c-infty}, assertion (iv) can also be proved directly, and then Lemma~\ref{lem:moore-osgood} applied to $f$, $g$, $h$ yields (iii). Thus the uniform estimate of Theorem~\ref{thm:uniform}, together with the computation for a fixed $\epsilon$ in Section~\ref{sec:nonrel}, provides a new proof of the non-relativistic limit \cite[Thm.~5.1]{HeTu22} as well.
\item Assumption~\ref{ass:v-prime} is needed only to identify the limit $L_\epsilon(z)$ in (ii) with the resolvent of the operator $H^\epsilon_A$. If $v$ satisfies Assumption~\ref{ass:v} and $v\notin H^1(\R)$, then all assertions of Theorem~\ref{thm:main} remain valid, with the same proof, if $H^\epsilon_A$ is replaced by the operator $T_\epsilon$ from Proposition~\ref{prop:Teps}; in this case no exceptional values of $\epsilon$ occur.
\end{enumerate}
\end{remark}

\subsection*{Renormalization of coupling constants}

Along the upper branch of the diagram in Theorem~\ref{thm:main}, no renormalization of coupling constants occurs: the approximating operators $D^\epsilon_{A,c}$ are obtained from the formal expression $\D_{A,c}$ by simply replacing $\ket{\delta}\bra{\delta}$ by $\ket{v_\epsilon}\bra{v_\epsilon}$ \cite{HeTu22}. On the other hand, the operator $H^\epsilon_A$ is \emph{not} obtained from \eqref{eq:formal-H} by replacing $\delta$, $\delta'$ by $u_1^\epsilon$, $u_2^\epsilon$; the coupling matrix $B_\epsilon$ differs from $A$. As $\epsilon\to0+$ we have
\[
B_\epsilon\to\frac{\det A}{a_{22}}E_{11}\quad\text{if}\ a_{22}\neq0,\qquad B_\epsilon = \begin{pmatrix}a_{11}+n_\epsilon a_{12}a_{21} & a_{12}\\ a_{21} & 0\end{pmatrix}\quad\text{if}\ a_{22}=0.
\]
In particular, $B_\epsilon = A$ for all $\epsilon$ if and only if $a_{22}=0$ and $a_{12}a_{21}=0$. To make the renormalization effect more explicit, we describe the limits of the ``naive'' approximations.

\begin{prop}\label{prop:naive}
Let $v$ satisfy Assumption~\ref{ass:v-prime}, let $B=(b_{jk})\in\C^{2,2}$ and put
\[
\widehat H^\epsilon_B := H_0+\sum_{j,k=1}^2 b_{jk}\ket{u_j^\epsilon}\bra{u_k^\epsilon},\qquad\Dom\widehat H^\epsilon_B = H^2(\R).
\]
\begin{enumerate}[(a)]
\item If $1+n_\epsilon b_{22}\neq0$, then $\widehat H^\epsilon_B = H^\epsilon_{A(\epsilon)}$ with $A(\epsilon) := B(I+n_\epsilon E_{22}B)^{-1}$.
\item If $b_{22}=0$ and $b_{12}b_{21}=0$, then $A(\epsilon) = B$ and $\widehat H_B^\epsilon\to H_B$ in the norm resolvent sense.
\item If $b_{22}\neq0$, then $A(\epsilon)\to A_* := \frac{\det B}{b_{22}}E_{11}$ as $\epsilon\to0+$, and $\widehat H^\epsilon_B$ converges in the norm resolvent sense to the $\delta$-interaction $H_{A_*}$ of strength $\det B/b_{22}$. More precisely, for every $z\in\res(H_{A_*})\setminus[0,+\infty)$ we have $\|(\widehat H^\epsilon_B-z)^{-1}-(H_{A_*}-z)^{-1}\|\le C(\theta(\epsilon)+\eta(\epsilon)+\epsilon)$ for all sufficiently small $\epsilon>0$. In particular, if $B=\lambda E_{22}$ with $\lambda\neq0$, then $\widehat H^\epsilon_B\to H_0$ in the norm resolvent sense.
\end{enumerate}
\end{prop}

\begin{proof}
Put $n:=n_\epsilon$. A straightforward computation shows that, if $1+nb_{22}\neq0$,
\begin{equation}\label{eq:Aeps-naive}
A(\epsilon) = B(I+nE_{22}B)^{-1} = \frac{1}{1+n b_{22}}\begin{pmatrix} b_{11}+n\det B & b_{12}\\ b_{21} & b_{22}\end{pmatrix}.
\end{equation}

(a) Write $A:=A(\epsilon)$. By \eqref{eq:Aeps-naive}, $a_{22} = b_{22}/(1+nb_{22})$, hence $1-na_{22} = (1+nb_{22})^{-1}\neq0$, and so $\det(I-nAE_{22}) = 1-na_{22}\neq0$. From $A(I+nE_{22}B) = B$ we get $(I-nAE_{22})B = A$, i.e., $B = (I-nAE_{22})^{-1}A$. Thus $B$ is the matrix \eqref{eq:Beps} associated with $A(\epsilon)$, and $\widehat H^\epsilon_B = H^\epsilon_{A(\epsilon)}$.

(b) If $b_{22}=0$, then $\det B = -b_{12}b_{21} = 0$ and \eqref{eq:Aeps-naive} gives $A(\epsilon)=B$. The convergence follows from Theorem~\ref{thm:main}(iv).

(c) Since $n_\epsilon\to+\infty$, \eqref{eq:Aeps-naive} yields $A(\epsilon)\to\frac{\det B}{b_{22}}E_{11} = A_*$.
Let $z\in\res(H_{A_*})\setminus[0,+\infty)$. By continuity of the determinant, $\det\big(I+\frac{i}{2}\alpha_1(\epsilon\sqrt z)A(\epsilon)\Zt(z)\big)\to\det(I+\frac{i}{2}A_*\Zt(z))\neq0$, so, for $\epsilon$ small, Theorem~\ref{thm:identification} and Theorem~\ref{thm:nonrel-eps} (computation of \eqref{eq:limit-kernel}) yield that $(\widehat H^\epsilon_B-z)^{-1}-K_0(z)$ is the operator with the kernel
\[
\frac14\Big[\Pi_\infty(v_\epsilon*F_\infty)(x)N^{(\epsilon)}(v_\epsilon*\check F_\infty)(y)\Pi_\infty\Big]_{11},
\]
where
\[
N^{(\epsilon)} := \Delta_\infty W\Big(I+\frac{i}{2}\alpha_1(\epsilon\sqrt z)A(\epsilon)\Zt(z)\Big)^{-1}A(\epsilon)W^*\Delta_\infty,
\]
cf.~Lemma~\ref{lem:c-infinity}(b). By Lemma~\ref{lem:c-infinity}(c), $(H_{A_*}-z)^{-1}-K_0(z)$ has the same form with $v_\epsilon*F_\infty$, $v_\epsilon*\check F_\infty$ and $N^{(\epsilon)}$ replaced by $F_\infty$, $\check F_\infty$ and $N_* := \Delta_\infty W(I+\frac{i}{2}A_*\Zt(z))^{-1}A_*W^*\Delta_\infty$, respectively. By \eqref{eq:Aeps-naive}, $|A(\epsilon)-A_*| = O(n_\epsilon^{-1}) = O(\epsilon)$, and $|\alpha_1(\epsilon\sqrt z)-1|\le\eta(\epsilon)$ by Lemma~\ref{lem:uniform-bounds}(d). Hence $|N^{(\epsilon)}-N_*| = O(\eta(\epsilon)+\epsilon)$. Moreover, $\|v_\epsilon*F_\infty-F_\infty\|_2$, $\|v_\epsilon*\check F_\infty-\check F_\infty\|_2\le\theta(\epsilon)$. The same decomposition as in the proof of Theorem~\ref{thm:uniform} shows that the Hilbert--Schmidt norm of the difference of the resolvents is $O(\theta(\epsilon)+\eta(\epsilon)+\epsilon)$. Finally, if $B=\lambda E_{22}$, then $\det B=0$, so $A_*=0$ and $H_{A_*}=H_0$.
\end{proof}

\begin{remark}
Proposition~\ref{prop:naive} shows that, in the non-relativistic setting, the non-local approximations obtained by the naive replacement $\delta\mapsto v_\epsilon$, $\delta'\mapsto v_\epsilon'$ in \eqref{eq:formal-H} reproduce only the $\delta$-interactions (and those interactions with $a_{22}=0$, $a_{12}a_{21}=0$); in all other cases the coupling constants have to be renormalized according to \eqref{eq:Beps}. This is in sharp contrast with the relativistic setting, where no renormalization is needed. The approximations $H_A^\epsilon$ obtained here as the non-relativistic limit of $D^\epsilon_{A,c}$ thus provide a natural explicit renormalization. A different approximation scheme, based on rank-one operators $\ket{v_\epsilon}\bra{u_\epsilon}$ with functions $u$ and $v$ having disjoint supports, which avoids the renormalization, was proposed in \cite{AlNi00}; see also \cite{AlNi07}.
\end{remark}

\section*{Acknowledgments}
The author was supported by the Czech Science Foundation (GA\v{C}R),
grant No.~26-21940S, and by the student grant project
SGS24/142/OHK4/3T/14 (CTU in Prague).
The author is grateful to Matěj Tušek for many valuable discussions.

\section*{AI disclosure}
During the preparation of this work the author used Claude Opus 5.5
(Anthropic), a large language model, for the following purposes:
language editing of the manuscript, suggesting useful notation, drafting parts of the
introduction and abstract, proposing the factorization of the resolvents in
Lemma~\ref{lem:factorization}, carrying out numerical checks of
several algebraic identities.

The author takes full responsibility for the content of the
publication.

\end{document}